\documentstyle[12pt]{article}

\newcommand{\bce}{\begin{center}}
\newcommand{\ece}{\end{center}}
\newcommand{\beq}{\begin{equation}}
\newcommand{\eeq}{\end{equation}}
\newcommand{\bea}{\vspace{0.25cm}\begin{eqnarray}}
\newcommand{\eea}{\end{eqnarray}}

\newcommand{\ba}{\begin{array}}
\newcommand{\ea}{\end{array}}

\def\lsim{\mathrel{\rlap{\lower4pt\hbox{\hskip1pt$\sim$}}
    \raise1pt\hbox{$<$}}}         
\def\gsim{\mathrel{\rlap{\lower4pt\hbox{\hskip1pt$\sim$}}
    \raise1pt\hbox{$>$}}}         

\def\Pom{{\bf I\!P}}

\def\lsim{\mathrel{\rlap{\lower4pt\hbox{\hskip1pt$\sim$}}
    \raise1pt\hbox{$<$}}}         
\def\gsim{\mathrel{\rlap{\lower4pt\hbox{\hskip1pt$\sim$}}
    \raise1pt\hbox{$>$}}}         

\def\Pom{{\bf I\!P}}

  \advance\oddsidemargin  by -0.9in
  \advance\evensidemargin by -0.9in
  \advance\topmargin      by -0.10in
\makeindex

\begin{document}

\begin{center}{\Large \bf
The diffraction cone for exclusive
vector meson production in deep inelastic
scattering}
\vspace*{0.5cm}\\
{\large \bf
J.~Nemchik$^{a,b}$,
N.N.~Nikolaev$^{c,d}$,
E.~Predazzi$^{a}$,
B.G.~Zakharov$^{d}$ and V.R.~Zoller$^{e}$
\bigskip\\}
{\it
$^{a}$ Dipartimento di Fisica Teorica, Universit\` a di Torino,\\
and INFN, Sezione di Torino, I-10125 Torino, Italy \medskip\\
$^{b}$ Institute of Experimental Physics, Slovak Academy of Sciences,\\
Watsonova 47, 04353 Ko\v sice, Slovak Republic \medskip\\
$^{c}$IKP(Theorie), KFA J{\"u}lich, 5170 J{\"u}lich, Germany \medskip\\
$^{d}$L. D. Landau Institute for Theoretical Physics, GSP-1,
117940, \\
ul. Kosygina 2, Moscow 117334, Russia\medskip\\
$^{e}$Institute for Theoretical and Experimental Physics,
B.Cheremushkinskaya 25, Moscow 117259, Russia}
\vspace{2.0cm}\\
\begin{minipage}[h]{13cm}
\centerline{\Large \bf
Abstract }
\vspace*{0.5cm}

We develop the 
color dipole gBFKL phenomenology of a
diffraction cone for photo- and electroproduction
$\gamma^{*}N \rightarrow VN$ of heavy vector mesons
(charmonium \& bottonium) at HERA and in fixed target
experiments. We predict a substantial shrinkage of the
diffraction cone
from the CERN/FNAL to the HERA range of c.m.s. energy $W$. The
$Q^{2}$-controlled selectivity to the color dipole size
(scanning phenomenon) is shown to lead to a decrease of the
diffraction slope with $Q^{2}$ (which is supported by the
available experimental data). We predict an approximate
flavor independence of the diffraction slope in the scaling
variable $Q^{2}+m_{V}^{2}$. For diffractive production of
the radially excited $2S$ states ($\Psi',\Upsilon'$) the
counterintuitive inequality of diffraction slopes
$B(2S) \lsim B(1S)$ is predicted, which defies the common wisdom
that diffraction slopes are larger for reactions with larger
size particles. 

\end{minipage}
\end{center}
\pagebreak
\setlength{\baselineskip}{0.70cm}

%
%
%
%
 
\section{Introduction}

Diffractive real and virtual photoproduction of ground state $V(1S)$
and radially excited $V'(2S)$ vector mesons
%
%
\beq
\gamma^{*}p \rightarrow V(1S)p,V'(2S)p
\label{eq:1.1}
\eeq
%
%
at high c.m.s. energy $W=\sqrt{s}$ is  an ideal testing ground
of ideas on the QCD pomeron exchange.
The new experimental data on vector meson production coming
from the HERA and fixed target experiments give a unique
insight into how the pomeron exchange evolves from the
nonperturbative to semiperturbative to perturbative regimes with
the increasing virtuality of the photon $Q^{2}$ and/or increasing
mass  $m_{V}$ of the produced vector meson and have prompted
intense theoretical discussions
\cite{KZ91,Ryskin,KNNZ93,KNNZ94,NNZscan,Brodsky,Forshaw,Ivanov,NNPZdipole,NNPZ96,Dosch}.

The usual approach to the perturbative QCD
(pQCD) pomeron is based on the BFKL equation \cite{Kuraev,Lipatov}
for the leading-log$s$ (LL$s$) evolution of the gluon distribution,
formulated in the scaling approximation of fixed QCD coupling
$\alpha_{S}=const$ and of infinite gluon correlation (propagation)
radius $R_{c}$ ( massless gluons); it sums the ladder
diagrams with reggeized $t$-channel gluon exchanges. 
More recently, a novel $s$-channel
approach to the LL$s$ BFKL equation
has been developed \cite{NZ94,NZZ94} in terms of the color dipole cross
section $\sigma(\xi,r)$ (hereafter  $r$ is the
color dipole moment, $x_{eff}= (m_{V}^{2}+Q^{2})/(W^{2}+Q^{2})$
and $\xi = log {1\over x_{eff}}$
is the rapidity variable). The color dipole approach, to be referred 
to as the running gBFKL approach, is favored 
because it incorporates consistently the two crucial properties
of QCD: i) asymptotic freedom (AF), {\sl i.e.},  the running 
QCD coupling $\alpha_{S}(r)$ and, ii) the finite
propagation radius $R_{c}$ of perturbative gluons.
AF and the running $\alpha_{S}(r)$ are an indispensable feature
of the
modern theory of deep inelastic scattering (DIS); without running
$\alpha_{S}(r)$ it is impossible to match  the leading-Log$Q^{2}$ 
(LL$Q^{2}$) limit of
the gBFKL equation with the conventional GLDAP equation \cite{GLDAP}
in the overlapping applicability region of the moderately
small $x$ (\cite{NZ94,NZBFKL}, see also \cite{Lipatov,Levin,Ross}).
The finite $R_{c}$ is of great importance
since the nonperturbative fluctuations in the QCD vacuum
restrict the phase space for the soft perturbative (real and
virtual) gluons and there is a strong evidence for finite $R_{c}$
from the lattice QCD studies (for the review see \cite{lattice})
and hadronic interactions \cite{Gotsman,Migdal}.
In the infrared region, one also needs to freeze $\alpha_{S}(r)$
in order not to run into the Landau pole,
$\alpha_{S}(r)\leq \alpha_{S}^{(fr)}$. Of course, if in our running 
gBFKL equation \cite{NZ94,NZZ94} one puts $R_{c}=\infty$  and 
$\alpha_{S}=const$, then the original scaling BFKL equation 
\cite{Kuraev,Lipatov} is recovered \cite{NZBFKL,Mueller}.

Being formulated in terms of real (rather
than reggeized) gluon exchanges, 
the color dipole running 
gBFKL equation \cite{NZ94,NZZ94} readily incorporates
the running $\alpha_{S}(r)$. The effect of finite $R_{c}$ can
be including by modifying the gluon propagator in the
infrared region, for instance, introducing the effective gluon 
mass, $\mu_{g}\approx 1/R_{c}$.
Remarkably, in LL$s$ approximation, a finite
$R_{c}$ is consistent with QCD gauge invariance.
The freezing of $\alpha_{S}(r)$ and the gluon correlation radius
$R_{c}$ are the nonperturbative parameters which describe the
transition from the soft, infrared, to the perturbative, hard,
region. The purely perturbative pomeron exchange
does not exhaust the scattering amplitude and in the practical
phenomenology of deep inelastic scattering one must add
certain soft nonperturbative exchange. It is important that 
the color dipole picture and color dipole factorization for
the proton structure function and for 
exclusive diffractive amplitudes do not require the validity of
pQCD and are viable also for the
soft pomeron exchange. The soft pomeron
exchange is important only for sufficiently large color dipoles,
$r > R_{c}$, and can readily be extracted from the
experimental data on hadronic cross sections \cite{NZ91} and
diffractive leptoproduction of light vector mesons
\cite{NNPZdipole}. On the
other hand, the pQCD, or hard, pomeron exchange can be related to the
perturbative
gluon structure function of the proton \cite{NZ94,BaroneSigma}.

Diffractive  production of $V(1S)$ mesons is particularly interesting
because of the so-called {\it scanning phenomenon}, by which the
production amplitude probes the color
dipole cross section at the dipole size $r\sim r_{S}$, where
%
%
\beq
r_{S} \approx {A \over \sqrt{m_{V}^{2}+Q^{2}}}\, ,
\label{eq:1.2}
\eeq
%
%
is the scanning radius \cite{KNNZ93,KNNZ94,NNZscan}.
This scanning property follows from the color dipole factorization
for production amplitudes and the shrinkage of the
transverse size of the virtual photon with $Q^{2}$  and
holds beyond the pQCD domain \cite{NZ91}.
Varying $Q^{2}$ one can study the transition from large,
nonperturbative and semiperturbative, dipole size $r_{S}$ to
the perturbative region of very short $r_{S}\ll R_{c}$ in
a very well controlled fashion \cite{NNPZdipole,NNPZ96}.
Furthermore, the scanning radius $r_{S}$ defines the transverse
size of the $\gamma^{*} V$ transition vertex, which contributes
to the total interaction radius and to the diffraction slope
$B(\gamma^{*}\rightarrow V)$.
Motivated by a remarkable phenomenological success of such a unified color 
dipole picture of hard and soft pomeron in application to the proton 
structure function \cite{NZHERA,NZZRegge} and vector meson production
\cite{NNZscan,NNPZ96}, in this communication we develop the color
dipole description of the forward diffraction cone 
$B(\gamma^{*}\rightarrow V)$ in exclusive diffractive DIS.
We use our early results for the energy dependence of the forward 
cone in color dipole scattering \cite{NZZslope}, obtained from the
solution of the running gBFKL equation for the diffraction slope
\cite{NZZspectrum}. Here the crucial point is that breaking of the scale
invariance by AF, i.e., by running  $\alpha_{S}(r)$, and finite $R_{c}$,
alters dramatically the very nature of the BFKL pomeron from a 
{\sl fixed cut} in the scaling approximation to a series of moving 
poles for the running gBFKL pomeron \cite{NZZspectrum,NZZRegge} (for early
quasiclassical analysis see also \cite{Lipatov}). 
As a result, in the running gBFKL approach
one predicts a substantial Regge shrinkage of the diffraction slope 
in the vector meson production, which can be tested at HERA.
In this paper we present for the first time the detailed analysis
of the $Q^{2}$ dependence and of the Regge growth of the diffraction
slope for the production of charmonium and bottonium states.

Diffractive production of radially excited  $V'(2S)$ mesons will
give an additional insight into the dipole size dependence of the
diffraction slope. Because of the node in the radial wave function of
the $V'(2S)$ states, there is a strong cancellation
between contributions to the production amplitude
from dipoles $r$ larger than or smaller than 
the node position $r_{n}$ (the node effect
\cite{KZ91,KNNZ93,NNZanom}). The resulting strong
suppression of diffractive production of $V'(2S)$ vs. $V(1S)$ 
has been confirmed experimentally 
in the $J/\Psi$ and $\Psi'$ photoproduction at HERA and in fixed target
experiments \cite{NMCPsi',E687Psi',H1Psi'}.
It has also interesting manifestations in the
differential cross sections, which we discuss in the present
paper for the first time. Because the radius of the $V'(2S)$
state is larger than the radius of the ground state $V(1S)$,
for the diffraction slopes one would naively expect the inequality
$B(\gamma^{*}\rightarrow \Psi') > B(\gamma^{*}\rightarrow J/\Psi)$.
However,
in this paper we demonstrate that the node effect in conjunction with
the color dipole factorization predicts a counterintuitive
inequality $B(\gamma^{*}\rightarrow \Psi') \lsim
B(\gamma^{*}\rightarrow J/\Psi)$, which can be tested at HERA.
Because the node effect is sensitive to the form of
the dipole cross section and its variation with energy,
we predict specific energy dependence of the $V'(2S)/V(1S)$
production ratios, which also can be tested at HERA.

The presentation is organized  as follows. The subject
of the introductory section 2 is the color dipole factorization
and the determination of the pQCD factorization scales
for diffractive production.
The running gBFKL formalism for the calculation
of the color dipole scattering matrix and of the diffraction
slope is presented in section 3. The decomposition of the 
diffraction slope into the perturbative
and nonperturbative components and on what physics controls the
$W^{2}$, flavor and $Q^{2}$ dependence of the diffraction slope
is expounded in section 4.
In Section 5 we discuss in more detail properties of the soft pomeron
exchange in the color dipole representation. 
In section 6 we present the salient features of the soft and hard exchanges on an example of $VN$ total cross sections. Predictions from
the running gBFKL dynamics for the forward and $t$-integrated 
vector mesons production cross
section are reported in section 7. We find a good agreement with
the low energy data and the data from the HERA collider experiments.
The subject of sections 8 is predictions for forward cone in diffractive
production of $V(1S)$ states with special emphasis on flavor symmetry.
Section 9 is concerned with the node effect in forward production 
of $V(2S)$ states.
The summary and some conclusions are presented in section 10.

%
%
%
%

\section{Introduction into color dipole factorization and
pQCD factorization scales for diffractive amplitudes}

The Fock state expansion for the
lightcone meson starts with the $q\bar{q}$ state; the snapshot
of the relativistic meson as a $q\bar{q}$ color dipole. The
probability amplitude to find the $q\bar{q}$ color dipole $\vec{r}$
is precisely the $q\bar{q}$ wave function, $\Psi({\vec{r}},z)$,
where $z$ is the fraction of meson's lightcone momentum carried
by a quark (the Sudakov lightcone variable). The interaction of
the relativistic color dipole of the dipole moment ${\vec{r}}$
with the target nucleon is quantified by the energy dependent color
dipole cross section, $\sigma(\xi,r)$. The effect of higher Fock
states $q\bar{q}g...$ is very important at high energy $\nu$.
To the LL$s$ and/or LL${1\over x}$ approximation it can be
reabsorbed into the energy (rapidity) dependence of $\sigma(\xi,r)$,
which is described by the running gBFKL equation \cite{NZ94,NZZ94}.
The dipole cross section is flavor independent and provides a
unified description of various diffractive processes.

In the limit of high photon energy $\nu$,  the $q\bar{q}$-nucleon 
the scattering matrix $\hat{\cal M}$
becomes diagonal in the mixed $(\vec{r},z)$-representation. This
$(\vec{r},z)$ diagonalization derives from the large longitudinal
coherence length,
%
%
\beq
l_{coh}\sim \frac{2\nu}{Q^{2}+m_{V}^{2}}\,,
\label{eq:2.1}
\eeq
%
%
and holds if
%
%
$
l_{coh}\gg R_{p}\, ,
$
%
%
where $R_{p}$ is a size of the target proton.
Because the coherence length is a purely kinematical scale \cite{Gribov},
the $(\vec{r},z)$
diagonalization does not require the  applicability of pQCD and 
must hold also for soft
pomeron exchange, i.e. even if the dipole size $\vec{r}$ is large. The
necessary condition is that the longitudinal scale $l_{soft}$
for the soft pomeron exchange is small, $l_{soft} \ll l_{coh}$,
which is the case for instance in the dual parton string model 
\cite{Veneziano} or different models of exchange by 
nonperturbative gluons \cite{Nachtmann,DL,Dosch}.
For the phenomenological success of a unified 
color dipole picture of vector meson production see \cite{NNPZdipole,NNPZ96}.

Taking advantage of the $(\vec{r},z)$ diagonalization of the scattering
matrix $\hat{\cal M}$, the amplitude for real (virtual) photoproduction
of vector mesons with the momentum transfer $\vec{q}$ can be
represented in the color dipole factorized form
%
%
\beq
{\cal M}(\gamma^{*}\rightarrow V,\xi,Q^{2},\vec{q})=
\langle V |\hat{\cal M}|\gamma^{*}\rangle=
\int\limits_{0}^{1} dz\int d^{2}\vec{r}
\Psi_{V}^{*}(r,z)
{\cal M}(\xi,r,z,\vec{q})
\Psi_{\gamma^{*}}(r,z)\,.
\label{eq:2.2}
\eeq
%
%
Our normalization is such that
%
%
$
\left.{d\sigma/ dt}\right|_{t=0}={|{\cal M}|^{2}/ 16\pi}.
$
%
%
In Eq. (\ref{eq:2.2}), $\Psi_{\gamma^{*}}(\vec{r},z)$ and
$\Psi_{V}(\vec{r},z)$ represent the probability amplitudes
to find the color dipole of size, $r$,
in the photon and quarkonium (vector meson), respectively
(for the sake of brevity we suppress the spin indices),
and ${\cal M}(\xi,r,z,\vec{q})$ is an amplitude for
elastic scattering of the color dipole on the target nucleon.
The color dipole distribution in (virtual) photons was
derived in \cite{NZ91,NZ94}.

The color dipole cross section $\sigma(\xi,r)$ only depends
on the dipole size $r$ but not on the $q$-$\bar{q}$ momentum
partition $z$. Because in the nonrelativistic heavy quarkonium $z\approx
\frac{1}{2}$, at small $\vec{q}$ in the diffraction cone 
one can safely neglect
the $z$-dependence of $\hat{\cal M}$ and set $z=\frac{1}{2}$.
Hereafter we will suppress the argument $z$.
Hereafter either $\xi$ or $x_{eff}$
resp. $x$
($x$ is Bjorken variable for the
inclusive DIS, the straightforward analysis of the relevant Sudakov
variables gives the relationship $x_{eff} \approx 2x$)
will be used whenever convenient.

We focus on calculating the imaginary part of the scattering
amplitude for which there is a simple representation in terms of
the gluon density matrix (see below). The small real part can easily be
reconstructed from analyticity considerations \cite{GribMig}
%
%
\beq
{\rm Re}{\cal M}(\xi,r,\vec{q}) =\frac{\pi}{2}\cdot\frac{\partial}
{\partial\xi} {\rm Im}{\cal M}(\xi,r,\vec{q})\,.
\label{eq:2.3}
\eeq
%
%
We suppress the discussion of ${\rm Re {\cal M}}$,
which is consistently included in all numerical results.

The details of calculation of the diffractive amplitude have been
presented elsewhere \cite{NNZscan,NNPZ96}. For the $Vq\bar{q}$
vertex function we assume the Lorentz structure $\Gamma\bar{\Psi}
\gamma_{\mu}\Psi V_{\mu}$.
For the $s$-channel helicity conservation  at small $\vec{q}$,
transverse (T) photons produce the transversely polarized vector
mesons and the longitudinally polarized (L) photons (to be more
precise, scalar photons) produce longitudinally polarized
vector mesons. One finds 
%
%
\bea
{\rm Im}{\cal M}_{T}(x_{eff},Q^{2},\vec{q})=
{N_{c}C_{V}\sqrt{4\pi\alpha_{em}} \over (2\pi)^{2}}
\cdot~~~~~~~~~~~~~~~~~~~~~~~~~~~~~~~~~
\nonumber \\
\cdot \int d^{2}{\bf{r}} \sigma(x_{eff},r,\vec{q})
\int_{0}^{1}{dz \over z(1-z)}\left\{
m_{q}^{2}
K_{0}(\varepsilon r)
\phi(r,z)-
[z^{2}+(1-z)^{2}]\varepsilon K_{1}(\varepsilon r)\partial_{r}
\phi(r,z)\right\}
\label{eq:2.4}
\eea
%
%
%
%
\bea
{\rm Im}{\cal M}_{L}(x_{eff},Q^{2},\vec{q})=
{N_{c}C_{V}\sqrt{4\pi\alpha_{em}} \over (2\pi)^{2}}
{2\sqrt{Q^{2}} \over m_{V}}
\cdot~~~~~~~~~~~~~~~~~~~~~~~~~~~~~~~~~
 \nonumber \\
\cdot \int d^{2}{\bf{r}} \sigma(x_{eff},r,\vec{q})
\int_{0}^{1}dz \,K_{0}(\varepsilon r)\left\{
[m_{q}^{2}+z(1-z)m_{V}^{2}]
\phi(r,z)-\partial_{r}^{2}
\phi(r,z)\right\}
\label{eq:2.5}
\eea
%
%
where
%
%
\beq
\varepsilon^{2} = m_{q}^{2}+z(1-z)Q^{2}\,,
\label{eq:2.6}
\eeq
%
%
$\alpha_{em}$ is the fine structure constant, $N_{c}=3$ is the number
of colors, $C_{V}={1\over \sqrt{2}},\,{1\over 3\sqrt{2}},\,{1\over 3},\,
{2\over 3},\,{1\over 3}~~$ are the charge-isospin factors for the
$\rho^{0},\,\omega^{0},\,\phi^{0},\, J/\Psi, \Upsilon$ production,
respectively and $K_{0,1}(x)$ are the modified Bessel functions.
The detailed discussion and parameterization of the lightcone radial
wave function $\phi(r,z)$ of the $q\bar{q}$ Fock state of the vector
meson is given in \cite{NNPZ96}. For heavy quarkonia one can safely
identify the current and constituent quarks. The terms $\propto
K_{0}(\varepsilon r)\phi(r,z)$ and
$\propto \varepsilon K_{1}(\varepsilon r)\partial_{r}\phi(r,z)$
for (T), 
$K_{0}(\varepsilon r)\partial_{r}^{2}\phi(r,z)$ for (L)
correspond to the helicity conserving and helicity-flip
transitions in the $\gamma^{*}\rightarrow q\bar{q},
V\rightarrow q\bar{q}$ vertices, respectively. In the
nonrelativistic heavy quarkonia, the helicity flip transitions are
the relativistic corrections, which become important only at large $Q^{2}$. 
Eq.~(\ref{eq:2.5}) corrects
a slight mistake in the relativistic correction to the amplitude for
production of longitudinally polarized photons made in \cite{NNZscan}.
The numerical results of Ref.~\cite{NNZscan} for the $J/\Psi$ are
only marginally different from those to be reported in this paper.

The representation for $\sigma(x,r,\vec{q})$ in terms
of the gluon density matrix (see Fig.~1)
%
%
\bea
\sigma(x,r,\vec{q})=
{4\pi \over 3}\int {d^{2}\vec{k}\over k^{4}}
\, \alpha_{S}(\kappa^{2})
[J_{0}({1\over 2}qr)-J_{0}(kr)]
{\cal F}(x,\vec{k}+{1\over 2}\vec{q},-\vec{k}+{1\over 2}\vec{q})
\label{eq:2.7}
\eea
%
%
where $J_{0}(x)$ is the usual Bessel function.
AF dictates that at the gluon-color dipole vertex, the QCD running
coupling must be taken at the largest relevant virtuality,
%
%
$
\kappa^{2}={\rm min}\{\vec{k}^{2},{C^{2}r^{-2}}\})\,,
$
%
%
where $C\approx 1.5$ \cite{NZ91}
and ensures the numerically similar results of calculations
in both the mixed $(r,z)$ and the momentum representations. 
The gluon density matrix
${\cal F}(x,\vec{k}+{1\over 2}\vec{q},-\vec{k}+{1\over 2}\vec{q})$
is proportional to the imaginary part of the non-forward gluon-nucleon
scattering amplitude; at $\vec{q}=0$ it equals the unintegrated gluon
structure function of the nucleon
%
%
$
{\cal F}(x,\vec{k},-\vec{k})
=\partial G(x,k^{2})/\partial \log k^{2}\,.
$
%
%
Eq.~(\ref{eq:2.7}) generalizes to the non-forward case $\vec{q}\neq 0$
the formula \cite{BaroneSigma,NZ94} for the dipole cross section
%
%
\beq
\sigma(x_{eff},r,\vec{q}=0)=\sigma(x_{eff},r)=
\frac{\pi^{2} r^{2}}{3}\int \alpha_{S}(\kappa^{2})
\frac{d k^{2}}{k^{2}}
\frac{4[1-J_{0}(kr)]}{(kr)^{2}}
\frac{\partial G(x_{eff},k^{2})}{\partial \log k^{2}}\,.
\label{eq:2.8}
\eeq
%
%
Because the function $f(y)=4[1-J_{0}(y)]/y^{2}$ can qualitatively be
approximated by the step-function,
%
%
$
f(y)\approx \theta(A_{\sigma}-y)\,,
$
%
%
where $A_{\sigma} \approx 10$ \cite{NZglue}, for small $r\ll R_{c}$
one readily finds
%
%
\beq
\sigma(x,r) =
\frac{\pi^2}{3}r^2\alpha_s(r)G(x,q_{\sigma}^2) \, ,
\label{eq:2.9}
\eeq
%
%
where the gluon structure function enters at the pQCD factorization scale
$q_{\sigma}^2 \sim {A_{\sigma}\over r^2}$ \cite{NZ94,BaroneSigma,NZglue}. 
For
large dipoles,
$r\gsim R_{c}$, one can neglect $J_{0}(kr)$ in the integrand, and the
dipole cross section saturates,
%
%
\beq
\sigma(x_{eff},r\gsim R_{c})=
{4\pi^{2} \over 3}\int \alpha_{S}(k^{2})
\frac{d k^{2}}{k^{4}}
\frac{\partial G(x_{eff},k^{2})}{\partial \log k^{2}}\,.
\label{eq:2.10}
\eeq
%
%

Next, notice that the integrands of (\ref{eq:2.4}),(\ref{eq:2.5})
are smooth at small $r$ and vanish exponentially at $r > 1/\epsilon$
due to $K_{0,1}(\epsilon r)$.
Because of the behavior $\sigma(x,r)\propto r^{2}$ in
(\ref{eq:2.9}), the amplitudes (\ref{eq:2.4}),(\ref{eq:2.5}) are
dominated by the contribution from the dipole size $r\approx r_{S}$
given by Eq.~(\ref{eq:1.2}) - the scanning phenomenon 
\cite{KNNZ93,KNNZ94,NNZscan}.
The scanning property is best quantified in terms of the
weight functions $W_{T,L}(Q^{2},r^{2})$ defined by
%
%
\beq
{\cal M}_{T}(x_{eff},Q^{2},\vec{q})=
{C_{V} \over (m_{V}^{2}+Q^{2})^{2}}
\int {dr^{2} \over r^{2}} {\sigma(x_{eff},r,\vec{q}) \over r^{2}}
W_{T}(Q^{2},r^{2})\,,
\label{eq:2.11}
\eeq
%
%
%
%
\beq
{\cal M}_{L}(x_{eff},Q^{2},\vec{q})=
{C_{V} \over (m_{V}^{2}+Q^{2})^{2}}
{2\sqrt{Q^{2}} \over m_{V}}
\int {dr^{2} \over r^{2}} {\sigma(x_{eff},r,\vec{q}) \over r^{2}}
W_{L}(Q^{2},r^{2})\,,
\label{eq:2.12}
\eeq
%
%
where in a somewhat abbreviated form ($i=T,L$, for the exact integrands
see Eqs.~(\ref{eq:2.4}),(\ref{eq:2.5}))
%
%
\beq
W_{i}(Q^{2},r^{2})=
{\pi \over C_{V}}r^{4}(m_{V}^{2}+Q^{2})^{2}
\int\limits_{0}^{1} dz\Psi_{V_{i}}^{*}(r,z)\Psi_{\gamma^{*}_{i}}(r,z)\,,
\label{eq:2.13}
\eeq
%
%
For the $1S$ mesons to a good approximation the so-defined
$W_{T,L}(Q^{2},r^{2})$ are sharply peaked functions of a
natural variable $y=\log r^{2}(Q^{2}+m_{V}^{2})$. The
height and width of the peak in $y$-distribution do only weakly
vary with $Q^{2}$ and the flavor and the peak position defines the
scanning radius $r_{S}\approx A_{T,L}/\sqrt{Q^{2}+m_{V}^{2}}$.
Consequently, the leading twist terms in the
expansion over the relevant short-distance parameter
$r_{S}^{2} \propto 1/(Q^{2}+m_{V}^{2})$ are of the form
(here we consider $\vec{q}=0$)
%
%
\beq
{\rm Im}
{\cal M}_{T} \propto
{1 \over Q^2+m_{V}^{2}}\sigma(x_{eff},r_{S}) \propto
 {1\over (Q^{2}+m_{V}^{2})^{2}}G(x_{eff},q_{T}^{2})\, ,
\label{eq:2.14}
\eeq
%
%
%
%
\beq
{\rm Im}
{\cal M}_{L} \approx {\sqrt{Q^{2}}\over m_{V}}{\cal M}_{T} \propto
{\sqrt{Q^{2}}\over m_{V}}
 {1\over (Q^{2}+m_{V}^{2})^{2}}G(x_{eff},q_{L}^{2})
\, .
\label{eq:2.15}
\eeq
%
%
By virtue of (\ref{eq:2.9}), here 
the pQCD scale
%
%
$
q_{T,L}^{2}=\tau_{L,T}(Q^{2}+m_V^{2})\, ,
$
%
%
where the scale parameter $\tau_{T,L}$ can be estimated as
%
%
\beq
\tau_{T,L}\approx \frac{A_{\sigma}}{A^{2}_{T,L}}\,.
\label{eq:2.16}
\eeq
%
%
For the more direct evaluation of the pQCD factorization scales
$q^{2}_{T,L}$ it is convenient to substitute (\ref{eq:2.8}) into
(\ref{eq:2.11}),(\ref{eq:2.12}), which then take the form reminiscent
of the  $k$-factorization formulas for $F_{2}(x,Q^{2})$
\cite{NZ91,k-fact}: 
%
%
\beq
{\rm Im}~{\cal M}_{T}(x_{eff},Q^{2},\vec{q}=0)=
{C_{V}\alpha_{S}(Q^{2}+m_{V}^{2}) \over (m_{V}^{2}+Q^{2})^{2}}
\int {dk^{2} \over k^{2}} {\partial G(x_{eff},k^{2})\over
\partial \log k^{2}}
\Theta_{T}(Q^{2},k^{2})\,
\label{eq:2.23}
\eeq
%
%
%
%
\beq
{\rm Im} {\cal M}_{L}(x_{eff},Q^{2},\vec{q}=0)=
{C_{V}\alpha_{S}(Q^{2}+m_{V}^{2}) \over (m_{V}^{2}+Q^{2})^{2}}
{2\sqrt{Q^{2}} \over m_{V}}
\int {dk^{2} \over k^{2}} {\partial G(x_{eff},k^{2}))
\over \partial \log k^{2}}
\Theta_{L}(Q^{2},k^{2})\,
\label{eq:2.24}
\eeq
%
%
where
%
%
\beq
\Theta_{T,L}(Q^{2},k^{2})=
\frac{\pi^{2}}{3}
\int \frac{d r^{2}}{r^{2}}
\frac{\alpha_{S}(\kappa^{2})}
{\alpha_{S}(Q^{2}+m_{V}^{2})}
\frac{4[1-J_{0}(kr)]}{(kr)^{2}}~ W_{T,L}(Q^{2},r^{2})\,.
\label{eq:2.25}
\eeq
%
%
Because of properties of $f(y)$  and of the sharp
peaking of $W_{T,L}(Q^{2},r^{2})$ at $r\approx r_{S}$, the weight
functions $\Theta_{T,L}(Q^{2},k^{2})$ are similar to the step
function,
%
%
\beq 
\Theta_{T,L}(Q^{2},k^{2}) \propto \theta (q_{T,L}^{2} -k^{2})\, ,
\label{eq:2.26}
\eeq
%
%
and 
%
\beq
\int {dk^{2} \over k^{2}} {\partial G(x_{eff},k^{2}))
\over \partial \log k^{2}}
\Theta_{i}(Q^{2},k^{2}))=
G(x_{eff},q_{i}^{2})
\int {dk^{2} \over k^{2}}
\Theta_{i}(Q^{2},k^{2}))=
G(x_{eff},q_{i}^{2})I_{i}(Q^{2})\, ,
\label{eq:2.27}
\eeq
%
%
where the factors
%
%
\beq
I_{T,L}(Q^{2})=
\frac{\pi^{2}}{3}
\int \frac{d r^{2}}{r^{2}}
\frac{\alpha_{S}(\kappa^{2})}
{\alpha_{S}(Q^{2}+m_{V}^{2})}
W_{T,L}(Q^{2},r^{2})
\label{eq:2.28}
\eeq
%
%
exhibit only a marginal dependence on $Q^{2}$.

For small $Q^{2}$ the scale parameters $A_{T,L}$ are close to the
nonrelativistic estimate $A \sim 6$, which follows from $r_{S}=
3/\varepsilon$ with the nonrelativistic choice $z=\frac{1}{2}$.
In general, $A_{T,L} \geq 6$ and increase slowly with $Q^2$
\cite{NNZscan}; for heavy quarkonia $A_{T,L}(\Upsilon)\sim 6$
at $Q^{2}\le 100$\,GeV$^{2}$ and $A_{T,L}(J/\Psi)\sim 6$ at
$Q^{2}=0$ and $A_{T,L}(J/\Psi)\sim 7$ at $Q^{2}=100\,$GeV$^{2}$, which
shows the relativistic corrections in the charmonium
and bottonium electroproduction are small. The corollary of the
large scanning radius $r_{S}$ and large values of $A_{T,L}$ is a very small
scale factor $\tau_{T,L}$ in the pQCD factorization
scale \cite{NNZscan}: $\tau_{T,L}(J/\Psi)
\approx 0.20$, $\tau_{L}(\rho^{0}) \approx 0.15$ and
$\tau_{T}(\rho^{0}) \approx 0.07-0.10$ for $Q^{2} \sim (10-100)
$\,GeV$^{2}$,
which are substantially smaller than 
$\tau \approx 0.25$ suggested in \cite{Ryskin} and $\tau \approx 1$
suggested in \cite{Brodsky}. Consequently,  the moderate
values of $Q^{2}$ attainable at HERA do, at the best, correspond
to the nonperturbative and semiperturbative values of $q_{T,L}^{2}$,
the soft 
contribution to the vector meson production must be substantial
and one must be careful with the interpretation of the vector meson
production data in terms of the gluon structure function. 
The point is that at $Q^{2}\lsim m_{J/\Psi}^{2}$ the scanning radius
$r_{S}$ is comparable to the radius of the $J/\Psi$, has been
overlooked in \cite{Ryskin}  and the formulas of Ref. \cite{Ryskin}
for the $J/\Psi$ production amplitudes in terms of the $J/\Psi$ 
wave function at the origin are too crude. Strictly
speaking, Eqs. (\ref{eq:2.23}), (\ref{eq:2.24}) and (\ref{eq:2.27})
were derived for the hard pQCD exchange when $r_{S}
\lsim R_{c}$ and/or for for the perturbatively large  $q_{T,L}^{2}$.
However, because the color dipole factorization is true beyond pQCD,
one can extend (\ref{eq:2.8}) to the soft pomeron and regard
this relationship as an operational definition of the nonperturbative
gluon distribution in the proton. To the same extent, Eqs.~(\ref{eq:2.23}),
(\ref{eq:2.24}) and (\ref{eq:2.27}) can serve as a unique basis for
extracting the whole gluon distribution, perturbative plus
nonperturbative, at small $x$ from the experimental data on diffractive
vector meson electroproduction at HERA.

The dominance (\ref{eq:2.15}) of the longitudinal
amplitude at $Q^{2}\gsim m_{V}^{2}$ follows, as a matter of fact,
from electromagnetic gauge invariance and as such it is true in any
reasonable model of vector meson production, the familiar
vector dominance model (VDM) included. The $Q^{2}$ dependence of
${\cal M}_{T,L}$ differs drastically from the VDM prediction
${\cal M}_T(VDM) \propto {1\over (m_V^2+Q^2)}\sigma_{tot}(\rho N)$,
though: instead of $\sigma_{tot}(\rho N)$ in (\ref{eq:2.14}) one
has  $\sigma(x_{eff},r_{S}) \propto r_{S}^{2} \propto
1/(Q^{2}+m_{V}^{2})$.

%
%
%
%
%

\section{The diffraction cone in the color dipole gBFKL approach}

In the familiar impact-parameter representation for amplitude of 
elastic scattering of the color 
dipole 
%
%
\beq
{\rm Im} {\cal M}(\xi,r,\vec{q})=2\int d^{2}\vec{b}\,
\exp(-i\vec{q}\vec{b})\Gamma(\xi,\vec{r},\vec{b})\,,
\label{eq:3.1}
\eeq
%
%
the diffraction slope $B=\left.- 
2{d \log {\rm Im}{\cal M}/ dq^{2}}\right|_{q=0}$
%
equals
%
%
\beq
B(\xi,r)= {1\over 2}\langle \vec{b}\,^{2}\rangle =
\lambda(\xi,r)/\sigma(\xi,r)\,,
\label{eq:3.2}
\eeq
%
%
where
%
%
\beq
\lambda(\xi,r)=\int d^2\vec{b}~
\vec{b}\,^2~\Gamma(\xi,\vec{r,}\vec{b})\, .
\label{eq:3.3}
\eeq
%
%
Then, the generalization of the color dipole factorization
formula (\ref{eq:2.2}) to the diffraction slope of the
reaction $\gamma^{*}p\rightarrow Vp$ reads:
%
%
\beq
B(\gamma^{*}\rightarrow V,\xi,Q^{2})
{\rm Im} {\cal M}(\gamma^{*}\rightarrow V,\xi,Q^{2},\vec{q}=0)=
\int\limits_{0}^{1} dz\int d^{2}\vec{r}\lambda(\xi,r)
\Psi_{V}^{*}(r,z)\Psi_{\gamma^{*}}(r,z)\,.
\label{eq:3.4}
\eeq
%
%

We sketch here the running gBFKL equation \cite{NZZslope} for 
$\lambda(\xi,r)$. The running gBFKL equation
for the energy dependence of the color dipole cross section reads
\cite{NZ94,NZZ94}
%
%
\bea
{\partial \sigma(\xi,r) \over \partial \xi} ={\cal K}\otimes
\sigma(\xi,r)=~~~~~~~~~~~~~~~~~~~~~~~~~~~~~~~~~~~~~~~~
\nonumber\\ {3 \over 8\pi^{3}} \int d^{2}\vec{\rho}_{1}\,\,
\mu_{G}^{2}
\left|g_{S}(R_{1})
K_{1}(\mu_{G}\rho_{1}){\vec{\rho}_{1}\over \rho_{1}}
-g_{S}(R_{2})
K_{1}(\mu_{G}\rho_{2}){\vec{\rho}_{2} \over \rho_{2}}\right|^{2}
[\sigma(\xi,\rho_{1})+
\sigma(\xi,\rho_{2})-\sigma(\xi,r)]   \, \, .
\label{eq:3.5}
\eea
%
%
Here the kernel ${\cal K}$ is related to the wave function squared
of the color-singlet $q\bar{q}g$ state with the Weizs\"acker-Williams
(WW) soft gluon, in which $\vec{r}$ is the $\bar{q}$-$q$ separation
and $\vec{\rho}_{1,2}$ are the $q$-$g$ and $\bar{q}$-$g$ separations
in the two-dimensional impact parameter plane.
The quantity
$\vec{{\cal E}}(\vec{\rho})= \mu_{G}g_{S}(\rho)K_{1}(\mu_{G}\rho)
{\vec{\rho} \over \rho}= -g_{S}(\rho) \vec{\nabla}_{\rho}
K_{0}(\mu_{G}\rho)$, where $K_{\nu}(x)$ is the modified Bessel
function, describes a Yukawa screened transverse chromoelectric field
of the relativistic quark and
%
%
\beq
\mu_{G}^{2}
\left|g_{S}(R_{1})
K_{1}(\mu_{G}\rho_{1}){\vec{\rho}_{1}\over \rho_{1}}
-g_{S}(R_{2})
K_{1}(\mu_{G}\rho_{2}){\vec{\rho}_{2} \over \rho_{2}}\right|^{2}
=
|\vec{{\cal E}}(\vec{\rho}_{1}) - \vec{{\cal E}}(\vec{\rho}_{2})|^{2}
\label{eq:3.6}
\eeq
%
%
describes the flux (the modulus of the Poynting vector) of WW gluons in
the $q\bar{q}g$ state. The asymptotic freedom of QCD uniquely
prescribes the chromoelectric field be computed with the running QCD charge
$g_{S}(r)=\sqrt{4\pi \alpha_{S}(r)}$ taken at the
shortest relevant distance, $R_{i}={\rm min}\{r,\rho_{i}\}$ in the
$q\bar{q}g$ system. The particular combination of the three color
dipole cross sections,
%
%
\beq
\Delta \sigma(\rho_{1},\rho_{2},r)=
{9\over 8}[\sigma(\xi,\rho_{1})+\sigma(\xi,\rho_{2})-\sigma(\xi,r)]\, ,
\label{eq:3.7}
\eeq
%
%
which emerges in the r.h.s. of the gBFKL equation,
is precisely a change of the color dipole cross section
for the presence of the WW gluon \cite{NZ94} in the $q\bar{q}g$
state.

At short distances, $r,\rho_{1,2}\ll R_{c}=1/\mu_{G}$, the kernel
${\cal K}$ does not depend on the infrared cutoff $R_{c}$.  The
Yukawa cut off of the long range chromoelectric field which has
been used in Eqs. (\ref{eq:3.5},\ref{eq:3.6}) is the simplest
phenomenological option. To the LL${1\over x}$ approximation,
this cutoff is consistent with gauge invariance.  If
one sacrifices  AF putting $g_{S}=const$ and lifts the infrared
cutoff by letting $R_{c}\rightarrow \infty$, one recovers the
scale-invariant kernel ${\cal K}$. Both the finite
$R_{c}$ and running $\alpha_{S}$ break the scale invariance,
the detailed discussion of
consequences is found in \cite{NZZ94,NZBFKL,NZZRegge,NZZspectrum}.
The principal phenomenon is that because of the lack of strong
log$r^{2}$ ordering in the BFKL equation there is an
intrusion from hard scattering to the regime of soft interactions
and vice versa, and the effect of the soft region is especially
enhanced by AF.
In the numerical analysis \cite{NZZ94} an infrared freezing
$\alpha_{S}(q^{2}) \leq \alpha_{S}^{(fr)}=0.82$ has been imposed
on the three-flavor, one-loop $\alpha_{S}(q^{2})=
4\pi/[9\log(k^{2}/\Lambda^{2})]$ with $\Lambda=0.3$\,GeV. With
$R_{c}=0.27$\,fm, i.e., $\mu_{G}=0.75$\,GeV, we found $\Delta_{\Pom}
=0.4$ \cite{NZZ94}, the calculation of Regge trajectories of
subleading pomeron singularities is reported in \cite{NZZRegge},
the emerging succesful description of the proton strcuture function
at small $x$ is published in \cite{NZHERA,NZZRegge}.

In \cite{NZZspectrum} the gBFKL equation (\ref{eq:3.5}) has been
generalized to the profile function $\Gamma(\xi,\vec{r},\vec{b})$,
where the impact parameter $\vec{b}$ is defined with respect
to the center of the dipole:
%
%
\bea
{\partial \Gamma(\xi,\vec{r},\vec{b})\over \partial \xi} =
{\cal K}\otimes \Gamma(\xi,\vec{r},\vec{b})
={3 \over 8\pi^{3}} \int d^{2}\vec{\rho}_{1}\,\,
\mu_{G}^{2}
\left|g_{S}(R_{1})
K_{1}(\mu_{G}\rho_{1}){\vec{\rho}_{1}\over \rho_{1}}
-g_{S}(R_{2})
K_{1}(\mu_{G}\rho_{2}){\vec{\rho}_{2} \over \rho_{2}}\right|^{2}
\nonumber\\
\times
[\Gamma(\xi,\vec{\rho}_{1},\vec{b}+{1\over 2}\vec{\rho}_{2}) +
\Gamma(\xi,\vec{\rho}_{2},\vec{b}+{1\over 2}\vec{\rho}_{1}) -
\Gamma(\xi,\vec{r},\vec{b})]  \, .~~~~~~~~~~~~~
\label{eq:3.8}
\eea
%
%
The calculation of the impact parameter integral (\ref{eq:3.1})
reduces Eq.~(\ref{eq:3.8}) to Eq.~({\ref{eq:3.5}). The
calculation of the moment (\ref{eq:3.3}) leads to the integral
equation for $\lambda(\xi,r)$. It is convenient to separate
from the diffraction slope
$B(\xi,r)$ the purely geometrical term $\frac{r^{2}}{8}$
related to the elastic form factor of the color dipole of the
dipole moment $r$ and to discuss instead of $\lambda(\xi,r)$
the function
%
%
$
\eta(\xi,r)=\lambda(\xi,r)-{1\over 8}r^{2}\sigma(\xi,r)\,,
$
%
%
which satisfies the inhomogeneous integral equation
%
%
\bea
{\partial \eta(\xi,r) \over \partial \xi} =
{3 \over 8\pi^{3}} \int d^{2}\vec{\rho}_{1}\,\,
\mu_{G}^{2}
\left|g_{S}(R_{1})
K_{1}(\mu_{G}\rho_{1}){\vec{\rho}_{1}\over \rho_{1}}
-g_{S}(R_{2})
K_{1}(\mu_{G}\rho_{2}){\vec{\rho}_{2} \over \rho_{2}}\right|^{2}
\nonumber\\
\times\left\{\eta(\xi,\rho_{1})+
\eta(\xi,\rho_{2})-\eta(\xi,r)+
{1\over 8}(\rho_{1}^{2}+\rho_{2}^{2}-r^{2})[
\sigma(\xi,\rho_2) + \sigma(\xi,\rho_1)]\right\}\nonumber\\
~~~~~~~~~~~~~~~~~~~~~~~
={\cal K}\otimes \eta(\xi,r) +\beta(\xi,r)
 \, ,
\label{eq:3.9}
\eea
%
%
where the inhomogeneous term equals
%
%
\bea
\beta(\xi,r) = {\cal L}\otimes \sigma(\xi,r)
={3 \over 64\pi^{3}} \int d^{2}\vec{\rho}_{1}\,\,
\mu_{G}^{2}
\left|g_{S}(R_{1})
K_{1}(\mu_{G}\rho_{1}){\vec{\rho}_{1}\over \rho_{1}}
-g_{S}(R_{2})
K_{1}(\mu_{G}\rho_{2}){\vec{\rho}_{2} \over \rho_{2}}
\right|^{2} \nonumber\\
\times(\rho_{1}^{2}+\rho_{2}^{2}-r^{2})[
\sigma(\xi,\rho_2) + \sigma(\xi,\rho_1)]
\, \, . ~~~~~~~~~~~~~~~~~~~~~~
\label{eq:3.10}
\eea
%
%

Beacuse the homogeneous piece of equation (\ref{eq:3.9}) coincides
with the gBFKL equation (\ref{eq:3.5}), asymptotically
the dipole cross section $\sigma(\xi,r)$ and the solution
$\eta(\xi,r)$ of the homogeneous Eq.~(\ref{eq:3.9}) have
identical energy dependence. Consequently, the solutions of the
homogeneous Eq.~(\ref{eq:3.9}) give the asymptotically
constant contribution to the diffraction cone and 
if  $\sigma_a(\xi,r)$ is a solution of Eq. (\ref{eq:3.5})
and $\eta_a(\xi,r)$ is a solution of
Eq. (\ref{eq:3.9}) with the diffraction slope $B_{a}(\xi,r)$,
then
%
%
$\eta_b(\xi,r) =
\eta_a(\xi,r)+\Delta{b}\cdot\sigma_a(\xi,r),
$
%
%
where $\Delta{b}=$const, is also a solution
of Eq. (\ref{eq:3.9}) with the diffraction slope
%
%
$
B_b(\xi,r) = B_a(\xi,r)+\Delta{b}\,.
$
%
%
It is the inhomogeneous term, $\beta(\xi,r)$, which  gives 
rise to $\eta(\xi,r) \propto \xi\sigma(\xi,r)$, i.e., to the 
asymptotic Regge growth of the diffraction
slope, $B(\xi,r)=B(\xi_{0},r)+2\alpha_{\Pom}'\xi$, and 
the Regge term $2\alpha'_{\Pom}\xi$ does not depend on
the size of the dipole $r$. Parametrically, $\alpha_{\Pom}' \propto
\alpha_{S}(R_{c}) R_{c}^{2}$ times a small numerical factor.
With the above specified infrared parameters  $\alpha_{\Pom}'
\approx 0.072$\,GeV$^{-2}$ was found in \cite{NZZslope}, for
slopes of subleading trajectories see \cite{NZZspectrum}.




\section{The beam, target and exchange decomposition of the
diffraction slope}

To have more insight into the dipole-size dependence of
the diffraction slope, it is useful to look at the scattering
amplitude $\sigma(\xi,r,\vec{q})$ in terms of the gluon density
matrix. For our purposes, it is sufficient to treat the
color structure of the proton in terms of the three valence 
(constituent) quarks.
Then, as illustrated graphically in Fig.~1b, the unintegrated
density matrix of gluons can be written
as
%
%
\bea
{\cal F}(x,\vec{k}+{1\over 2}\vec{q},-\vec{k}+{1\over 2}\vec{q})=
{4 \over \pi}
\int d^{2}\vec{k}_{1}
{\cal T}(\xi,\vec{k}+{1\over 2}\vec{q},
-\vec{k}+{1\over 2}\vec{q},\vec{k}_{1}+{1\over 2}\vec{q},
-\vec{k}_{1}+{1\over 2}\vec{q})\nonumber\\
\alpha_{S}(k_{1}^{2})
[G_{1}(q^{2})-G_{2}(\vec{k}_{1}+{1\over 2}\vec{q},
-\vec{k}_{1}+{1\over 2}\vec{q})]\,,
\label{eq:4.1}
\eea
%
%
where $G_{1}(q^2)$ and $G_{2}(\vec{\kappa}_{1},\vec{\kappa}_{2})$ are
the single- and two-quark form factors of the proton probed by
gluons and
${\cal T}(\xi,\vec{k}+\vec{q},
-\vec{k}+{1\over 2}\vec{q},\vec{k}_{1}+\vec{q},
-\vec{k}_{1}+{1\over 2}\vec{q})$
stands for the propagation function of two $t$-channel gluons.
In the Born approximation,
%
%
\beq
{\cal T}(\xi,\vec{k}+{1\over 2}\vec{q},
-\vec{k}+{1\over 2}\vec{q},\vec{k}_{1}+{1\over 2}\vec{q},
-\vec{k}_{1}+{1\over 2}\vec{q})
={\delta(\vec{k}-\vec{k}_{1})\over
[(\vec{k}+{1\over 2}\vec{q})^{2}+\mu_{G}^{2}]
[(\vec{k}-{1\over 2}\vec{q})^{2}+\mu_{G}^{2}]}\, .
\label{eq:4.2}
\eeq
%
%

Splitting the color dipole vertex function into two pieces,
$V_{d}(q,r)=[J_{0}({1\over 2}qr)-J_{0}(kr)]=
[J_{0}({1\over 2}qr)-1]+[1-J_{0}(kr)]$, we obtain a useful
decomposition
%
%
\bea
\sigma(\xi,r,\vec{q})=
{4\pi \over 3}[J_{0}({1\over 2}qr)-1]\int {d^{2}\vec{k}\over k^{4}}
\, \alpha_{S}(\kappa^{2})
{\cal F}(x,\vec{k}+{1\over 2}\vec{q},-\vec{k}+{1\over 2}\vec{q})
\nonumber\\
+{4\pi \over 3}\int {d^{2}\vec{k}\over k^{4}}
\, \alpha_{S}(\kappa^{2})
[J_{0}(kr)-1]
{\cal F}(x,\vec{k}+{1\over 2}\vec{q},-\vec{k}+{1\over 2}\vec{q})
\label{eq:4.3}
\eea
%
%
Because of the property (\ref{eq:2.12}), the second term has the
typical logarithmic $k^{2}$ integration.  It comprises the contributions
to the $q$ dependence from the target and exchanged gluons. In
contrast, such a logarithmic $k^{2}$ integration is absent in the
first term; here the  $k^{2}$ integration converges at
finite $k^{2}\sim R_{c}^{-2}$.

The emerging representation
%
%
\bea
\sigma(\xi,r,\vec{q})=
{4\pi \over 3}[J_{0}({1\over 2}qr)-1]\int {d^{2}\vec{k}\over k^{4}}
\, \alpha_{S}(\kappa^{2})
{\cal F}(x,\vec{k}+{1\over 2}\vec{q},-\vec{k}+{1\over 2}\vec{q})\nonumber\\
+
{16 \over 9}\int {d^{2}\vec{k}\over k^4}\, \alpha_{S}(\kappa^{2})
[1-J_{0}(kr)]\nonumber\\
\cdot
\int d^{2}\vec{k}_{1}
{\cal T}(\xi,\vec{k}+\vec{q},
-\vec{k}+{1\over 2}\vec{q},\vec{k}_{1}+{1\over 2}\vec{q},
-\vec{k}_{1}+{1\over 2}\vec{q})
\alpha_{S}(k_{1}^{2})
[G_{1}(q^{2})-G_{2}(\vec{k}_{1}+{1\over 2}\vec{q},
-\vec{k}_{1}+{1\over 2}\vec{q})] \, ,
\label{eq:4.4}
\eea
%
%
nicely illustrates how the
three relevant size parameters in the problem give
rise to the three major components of the diffraction slope.
The $q$ dependence coming from the proton vertex function
$V_{p}(\vec{k}_{1},\vec{q})=
G_{1}(q^{2})-G_{2}(\vec{k}_{1}+{1\over 2}\vec{q},
-\vec{k}_{1}+{1\over 2}\vec{q})$ is controlled by the proton
size.  The $q$ dependence coming from the color dipole vertex
function $V_{d}=J_{0}({1\over 2}qr)-1$ is controlled
by the color dipole size $r$. The $q$ dependence coming
from ${\cal T}(\xi,\vec{k}+{1\over 2}\vec{q},
-\vec{k}+{1\over 2}\vec{q},\vec{k}_{1}+{1\over 2}\vec{q},
-\vec{k}_{1}+{1\over 2}\vec{q})
$ depends on the effective $k^{2},k_{1}^{2}$ which
contribute to the scattering amplitude and on the gluon propagation
radius $R_{c}$. The latter scale remains important even at large
$k$ because  the properties of the running 
gBFKL pomeron
are controlled by interactions at $r\sim R_{c}$. In the 
asymptotic BFKL
regime, at small $x$, the $\vec{k_{1}}$ and $\vec{k}$ become azimuthally
uncorrelated.

In order to proceed further, one needs a model for $G_{1}(q^2)$
and $G_{2}(\vec{\kappa}_{1},\vec{\kappa}_{2})$. The radius of the proton
$R_{N}$ probed by the gluon can be different from the charge radius
$R_{ch}$, still $R_{ch}$ serves as a useful scale.
The two-quark form factor $G_{2}(\vec{k}+{1\over2}\vec{q},
-\vec{k}+{1\over 2}\vec{q})$ is a steep function of $k^{2}$
and a smoother function of $q^{2}$ \cite{BGZfactor}. For instance, for the
oscillator wave function of the 3-quark proton one readily finds
%
%
\beq
G_{2}(\vec{k}+{1\over 2}\vec{q},
-\vec{k}+{1\over 2}\vec{q})] =
G_{1}({1\over 4}q^{2})G_{1}(3k^{2})\, .
\label{eq:4.5}
\eeq
%
%

A straightforward differentiation gives a transparent decomposition
of $d\sigma(\xi,r,\vec{q})/dq^{2}$ into the following four terms
%
%
\bea
\left.{d\sigma(\xi,r,\vec{q})\over dq^{2}}\right|_{q^{2}=0}=
\sum_{i=1}^{4}\left.{d\sigma^{(i)}
(\xi,r,\vec{q})\over dq^{2}}\right|_{q^{2}=0}=
-{16\over 3}\int {d^{2}\vec{k}\over k^4}\, \alpha_{S}(\kappa^{2})
\nonumber\\
\left\{
{1\over 16}r^{2}\int d^{2}\vec{k}_{1}
\int d^{2}\vec{k}_{1}
{\cal T}(\xi,\vec{k},
-\vec{k},\vec{k}_{1},
-\vec{k}_{1})\alpha_{S}(k_{1}^{2})
[1-G_{2}(\vec{k}_{1},\vec{k}_{1})]
\right.
\nonumber\\
-
\left.[1-J_{0}(kr)]\int d^{2}\vec{k}_{1}
\alpha_{S}(k_{1}^{2})
[1-G_{2}(\vec{k}_{1},\vec{k}_{1})]
{\partial {\cal T}(\xi,\vec{k}+{1\over 2}\vec{q},
-\vec{k}+{1\over 2}\vec{q},\vec{k}_{1}+{1\over 2}\vec{q},
-\vec{k}_{1}+{1\over 2}\vec{q})\over \partial q^{2}}\right|_{q^{2}=0}
\nonumber\\
+
\left.{1\over 6}R_{N}^{2}[1-J_{0}(kr)]\int d^{2}\vec{k}_{1}
{\cal T}(\xi,\vec{k},
-\vec{k},\vec{k}_{1},
-\vec{k}_{1})\alpha_{S}(k_{1}^{2})
\right.
\nonumber\\
-
\left.{1\over 24}R_{N}^{2}[1-J_{0}(kr)]\int d^{2}\vec{k}_{1}
{\cal T}(\xi,\vec{k},
-\vec{k},\vec{k}_{1},
-\vec{k}_{1})\alpha_{S}(k_{1}^{2})
G_{2}(\vec{k}_{1},\vec{k}_{1})
\right\} 
\nonumber\\
\label{eq:4.6}
\eea
%
%
The following properties of ${\cal T}(\xi,\vec{k}+{1\over 2}\vec{q},
-\vec{k}+{1\over 2}\vec{q},\vec{k}_{1}+{1\over 2}\vec{q},
-\vec{k}_{1}+{1\over 2}\vec{q})$ are
important in (\ref{eq:4.6}): First, in the infrared regulated
QCD it is nonsingular at $k^{2}=0$, cf. Eq.~(\ref{eq:4.2}). Second,
(modulo to the logarithmic scaling violations) its large-$k^{2}$
asymptotic is similar to that of the Born term (\ref{eq:4.2}),
${\cal T}(\xi,\vec{k}+{1\over 2}\vec{q},
-\vec{k}+{1\over 2}\vec{q},\vec{k}_{1}+{1\over 2}\vec{q},
-\vec{k}_{1}+{1\over 2}\vec{q})\propto 1/k^{-4}\, .
$
Third, in the Born approximation (after the azimuthal averaging)
%
%
\bea
\left.
{\partial {\cal T}(\xi,\vec{k}+{1\over 2}\vec{q},
-\vec{k}+{1\over 2}\vec{q},\vec{k}_{1}+{1\over 2}\vec{q},
-\vec{k}_{1}+{1\over 2}\vec{q}) \over \partial q^{2}}\right|_{q^{2}=0}
=\nonumber\\
-{R_{c}^{2} \over (1+R_{c}^{2}k^{2})^{2}}
{\cal T}(\xi,\vec{k}+{1\over 2}\vec{q},
-\vec{k}+{1\over 2}\vec{q},\vec{k}_{1}+{1\over 2}\vec{q},
-\vec{k}_{1}+{1\over 2}\vec{q}) \, .
\label{eq:4.7}
\eea
%
%
Fourth, the finding of the asymptotic Regge growth of the diffraction
slope in \cite{NZZslope} implies that in the high energy limit
$\xi \rightarrow \infty$ and for all $\vec{k},\vec{k}_{1}$
%
%
\bea
\left.
{\partial {\cal T}(\xi,\vec{k}+{1\over 2}\vec{q},
-\vec{k}+{1\over 2}\vec{q},\vec{k}_{1}+{1\over 2}\vec{q},
-\vec{k}_{1}+{1\over 2}\vec{q}) \over \partial q^{2}}\right|_{q^{2}=0}
=\nonumber\\
-\left[\alpha_{\Pom}'(\xi-\xi_{0})+{\cal O}(R_{c}^{2})\right]
{\cal T}(\xi,\vec{k}+{1\over 2}\vec{q},
-\vec{k}+{1\over 2}\vec{q},\vec{k}_{1}+{1\over 2}\vec{q},
-\vec{k}_{1}+{1\over 2}\vec{q}) \, .
\label{eq:4.8}
\eea
%
%

Consider first the decomposition of the diffraction slope
for large dipoles, $r \gsim R_{c}$. In this
limit, the amplitude (\ref{eq:4.4}) is dominated by the contribution
from $k^{2}\sim \mu_{G}^{2}=R_{c}^{-2}\gg R_{N}^{-2}$, so that
$J_{0}(kr), G_{1}(3k^{2})\ll 1$ and can be neglected
and
%
%
$
\sigma(\xi,r)=
{4\pi \over 3}\int d^{2}\vec{k}\, \alpha_{S}^{2}(k^{2})
{\cal F}(\xi,\vec{k},\vec{k})\,,
$ cf. Eq.~(\ref{eq:2.7}).
%
%
Then, the first term in the expansion (\ref{eq:4.6}) can be
evaluated as
%
%
\beq
\left.{d\sigma^{(1)}(\xi,r,\vec{q})
\over dq^{2}}\right|_{q^{2}=0}=
-
{1\over 2}\Delta {b}_{1} \sigma(\xi,r)
=
-
{1\over 16}r^{2}\sigma(\xi,r)\, .
\label{eq:4.9}
\eeq
%
%
Similarly,
%
%
\beq
\left.{d\sigma^{(3)}(\xi,r,\vec{q})
\over dq^{2}}\right|_{q^{2}=0}=
-
{1\over 2}\Delta b_{3}\sigma(\xi,r)
=
-{1\over 6}R_{N}^{2}\sigma(\xi,r)\, .
\label{eq:4.10}
\eeq
%
%
The integrand of the fourth term contains the steeply decreasing
two-body form factor $G_{2}(\vec{k},-\vec{k})$, which cuts off
the integration at $k^{2} \lsim R_{N}^{2}$. Consequently, one must
distinguish between $r\lsim R_{N}$ and $r\gsim R_{N}$. A simple estimate,
which interpolates between these limiting cases, is
%
%
\beq
\left.{d\sigma^{(4)}(\xi,r,\vec{q})\over dq^{2}}\right|_{q^{2}=0}=
-{1\over 2}\Delta b_{4}\sigma(\xi,r)=
-{1\over 24}R_{c}^{2}\sigma(\xi,r){r^{2} \over r^{2}+R_{N}^{2}}\,.
\label{eq:4.11}
\eeq
%
%
The bottom line is that $\Delta b_{4}\ll \Delta b_{3}$.
Finally, making use of (\ref{eq:4.8}), the second term in (\ref{eq:4.6})
can be estimated as
%
%
\beq
\left.{d\sigma^{(2)}(\xi,r,\vec{q})\over dq^{2}}\right|_{q^{2}=0}=
-{1\over 2}\Delta b_{2}\sigma(\xi,r) =
-\left[\alpha_{\Pom}'(\xi-\xi_{0})+{\cal O}(R_{c}^{2})\right]
\sigma(\xi,r)\,.
\label{eq:4.12}
\eeq
%
%
At low energy, in the Born approximation, Eq.~(\ref{eq:4.7}) gives
$\Delta b_{2}=2R_{c}^{2}$. The salient feature of the
resulting diffraction slope
%
%
\beq
B(\xi,r)=\sum_{i} \Delta b_{i}=
\frac{1}{8}r^{2}+\frac{1}{3}R_{N}^{2}+
2\alpha_{\Pom}'(\xi-\xi_{0}) + {\cal O}(R_{c}^{2})\, ,
\label{eq:4.13}
\eeq
%
%
is a presence of the geometrical contributions
$\Delta b_{1}={1\over 8}r^{2}$ and $\Delta b_{3}={1\over 3}R_{N}^{2}$.

For large dipoles $r\gsim R_{c}$ one recovers a sort of
additive quark model, in which the uncorrelated gluonic clouds
build up around the beam and target quarks and antiquarks and
the terms ${\cal O}( R_{c}^{2})$ and $2\alpha_{\Pom}'(\xi-\xi_{0})$
describe the familiar Regge growth of diffraction slope for
the quark-quark scattering.
The opposite limit of small dipoles, $r\ll R_{c}$, is somewhat
more tricky.
In the second and third term, the $k^{2}$ integration is cut
off by $[1-J_{0}(kr)]$ and extends up to $A_{\sigma}/r^{2}$,
precisely as in the dipole cross section (\ref{eq:2.8}).
Consequently, their contributions to the derivative (\ref{eq:4.6})
are still given by Eq.~(\ref{eq:4.12}) and Eq.~(\ref{eq:4.10}),
respectively, so that the Regge term and the contribution from
the target proton size to expansion (\ref{eq:4.13}) are retained.
The contribution from the first term, i.e., from the size of the
color dipole, changes dramatically and will no longer have the
geometric form ${1\over 8}r^{2}$. Indeed, as we discussed following
Eq. (\ref{eq:4.3}), the $k^{2}$ integration in the first term in
(\ref{eq:4.6}) converges at $k^{2} \lsim R_{c}^{2}$. Consequently,
in this limit $\kappa^{2}=C^{2}/r^{2}$ and one can factor out
$\alpha_{S}(\kappa^{2})=\alpha_{S}(r)$ from the integrand.
This leads to an estimate
%
%
\bea
\left.{d\sigma^{(1)}(\xi,r,\vec{q})
\over dq^{2}}\right|_{q^{2}=0}\approx -
{r^{2}\over 16}\cdot {\alpha_{S}(r) \over \alpha_{S}(R_{c})}
\sigma(\xi,R_{c})
\approx -{r^{2} \over 16}\cdot {\pi^{2}\over 3} \alpha_{S}(r)
R_{c}^{2} G(\xi,{A_{\sigma}\over R_{c}^{2}})
\label{eq:4.14}
\eea
%
%
and, after making use of (\ref{eq:2.9}), to
%
%
\beq
\Delta b_{1} = {R_{c}^{2} \over 8}{G(\xi,{A_{\sigma}/ R_{c}^{2}})
\over G(\xi,{A_{\sigma}/r^{2}})}
\label{eq:3.28}
\eeq
%
%
Similar considerations give an estimate for the contribution
to the diffraction slope from the fourth term in (\ref{eq:4.6}),
which is a negligible correction to $\Delta b_{1}$:
%
%
\beq
\Delta b_{4} = {R_{c}^{2} \over 12}
\cdot {R_{c}^{2} \over R_{N}^{2}}\cdot
{G(\xi,{A_{\sigma}/ R_{c}^{2}})
\over G(\xi,{A_{\sigma}/r^{2}})}\,.
\label{eq:3.29}
\eeq
%
%

More comments on $\Delta b_{1}$ are in order. At asymptotically
large $\xi$ and/or asymptotically small $x$, the running gBFKL approach
predicts the universal $x$ dependence of the gluon structure
function \cite{NZBFKL}
%
%
\beq
G(x,Q^{2}) \propto \left[{1\over \alpha_{S}(Q^{2})}\right]^{\gamma}
\cdot \left({1 \over x}\right)^{\Delta_{\Pom}}
\label{eq:3.30}
\eeq
%
%
where $\gamma = {12\over \beta_{0}}\Delta_{\Pom}$ and $\beta_{0}=11-
{2\over 3}n_{f}$. Consequently, in the well developed BFKL regime
$\Delta b_{1}$ will not depend on energy:
%
%
\beq
\Delta b_{1} = {R_{c}^{2} \over 8}
\left[{\alpha_{S}(r)\over
\alpha_{S}(R_{c})}\right]^{\gamma}
\label{eq:3.31}
\eeq
%
%
However, at moderately small $x$ values, the $x$ dependence of the gluon
structure function exhibits strong dependence on the factorization
scale, the ratio ${G(\xi,{A_{\sigma}\over R_{c}^{2}})
/G(\xi,{A_{\sigma}\over r^{2}})}$ has a substantial $x$ dependence
and $\Delta b_{1}$ contributes to the energy dependence of the diffraction
cone. Specifically, it makes the slope of the effective Regge
trajecrory $\alpha'_{\Pom}(eff)$ substantially larger than
the true slope of the leading Pomeron trajectory $\alpha_{\Pom}'$
\cite{NZZslope}.

To summarize, the geometrical contribution to the diffraction
slope from the target proton size, $\Delta b_{3}={1\over 3}R_{N}^{2}$,
persists for all the dipole sizes (the term $\Delta b_{4}$ which
is also associated with the proton size is negligibly small in all
cases). Although the nonperturbative parameter $R_{N}^{2}$ is not
calculable from first principles, its contribution to the
diffraction slope varies neither with energy nor with the dipole size
and can eventually be fixed from the accurate experimental data.

%
%
%
%

\section{Soft pomeron and diffractive scattering of large
color dipoles}

The need for a soft pomeron contribution in addition to the
gBFKL dipole cross section described previously is brought about by
phenomenological considerations. 
A viable gBFKL phenomenology of the rising component of the proton
structure function over the whole range of $Q^{2}$ studied at
HERA (real photoabsorption included) is obtained if one starts
with the Born dipole cross section  $\sigma_{B}(r)$ 
as a boundary condition
for the gBFKL evolution at $x_{0}=0.03$ \cite{NZHERA,NZZRegge}.
However, such a $\sigma_{B}(r)$ falls short of the
interaction strength at $r\gsim R_{c}$; roughly speaking, for
the phenomenological value $R_{c} = 0.27$\,fm one finds
$\sigma_{B}(r\gsim 1\,{\rm fm})\sim 5$
\, mb, whereas for the description of soft processes  one
rather needs the dipole cross section $\sim 50$ mb at
$r\gsim 2$\,fm.
Therefore, at $r\gsim R_{c}$, the above described perturbative gBFKL
dipole cross section (which hereafter we supply with the
subscript ``pt'') $\sigma_{pt}(\xi,r)$, must be complemented
by the contribution from the nonperturbative soft pomeron,
$\sigma_{npt}(\xi,r)$. Because in all the cases studied
the contribution from $\sigma_{pt}(\xi,r)$ exhausts the
rise of the total cross sections and/or of the proton structure
function, in Refs.~\cite{NZHERA,NNZscan} we have modeled the
soft nonperturbative pomeron  by the energy independent
$\sigma_{npt}(\xi,r)=\sigma_{npt}(r)$.
For the lack of better theoretical and experimental information
as well as simplicity,
we make the simplest possible assumption that the eikonals for the
perturbative and soft interactions are additive, which to
the lowest order amounts to additivity of the dipole cross
sections $\sigma(\xi,r)=\sigma_{pt}(\xi,r)+\sigma_{npt}(r)$.

The direct determination of the total dipole cross section 
$\sigma(\xi,r)$,
from the experimental data on photo- and leptoproduction of
vector mesons is reported in \cite{NNPZdipole} and supports
the flavor independence of $\sigma(\xi,r)$. Other constraints
for $\sigma_{npt}(r)$ include real photoproduction
\cite{NNZscan,NNPZ96}, hadronic diffractive scattering \cite{NZ91},
nuclear shadowing in deep inelastic scattering \cite{BaroneShad},
diffractive deep inelastic scattering at HERA
\cite{GNZ95, NSZ96}, nuclear attenuation in
photoproduction of light vector mesons and the onset of
color transparency in leptoproduction of vector mesons
\cite{KNNZ94} and the the proton structure function at moderate
and small $Q^{2}$ \cite{NZHERA,NZZRegge}. All the results are
consistent with the form of the dipole cross section suggested in
\cite{NZ91,NZHERA,NNZscan}, a convenient parameterization for
which is
%
%
\beq
\sigma_{npt}(r)=
\sigma_{0}
\left[1-\sum\limits_{i=1}^{2}A_{i}
\exp(-{r^{2}\over a^{2}_{i}})\right]\cdot
\left[1+\sum\limits_{i=1}D_{i}\exp(-{(r-b_{i})^{2}\over c^{2}_{i}})
\right]\,.
\label{eq:5.1}
\eeq
%
%
with $\sigma_{0}=41.2$ mb, $A_{1}=1.45$, $A_{2}=-0.45$,
$a_{1}=1.30$ fm, $a_{2}=0.75$ fm, $D_{1}=0.80$, $D_{2}=0.36$,
$b_{1}=0.88$ fm, $b_{2}=2.08$ fm, $c_{1}=0.53$ fm, $c_{2}=1.14$ fm.
For a somewhat cruder fit
with $D_{i}=0$ we find $\sigma_{0}=51.6$ mb, $A_{1}=1.82$, $A_{2}=-0.82$,
$a_{1}=1.05$ fm, $a_{2}=0.72$ fm.
For small dipoles, $r\ll R_{c}$, this cross section is poorly
known because it is swamped by $\sigma_{pt}(\xi,r)$.

There isn't anything unusual in the concept of a nonperturbative
cross section. The conventional gluon structure function of the
photon
%
%
$$
G(x,Q^{2})=\int_{0}^{Q^{2}} {dk^{2} \over k^{2}}
{\cal F}(x,\vec{k},-\vec{k})
$$
%
%
always contains a contribution from gluons with soft transverse
momenta $k^{2} < Q_{0}^{2} \lsim $1\,GeV$^{2}$, which persists
at all $Q^{2}$ and equals precisely $G(x,Q_{0}^{2})$, the
familiar input to the conventional GLDAP analysis of the
$Q^{2}$ evolution of parton densities. One is perfectly 
content
with the strong sensitivity of the GLDAP evolution to this
unknown soft input $G(x,Q_{0}^{2})$, which is routinely fixed
from fits to the experimental data. In the color dipole approach
to DIS, our soft dipole cross section $\sigma_{npt}(r)$ plays
exactly the same r\^ole as the gluon (quark) structure functions
at a soft scale $Q_{0}^{2}$. Furthermore, it is 
tempting to reinterpret this soft dipole cross
section $\sigma_{npt}(r)$ in terms of the nonperturbative gluon
distribution in the spirit of Eq.~(\ref{eq:2.8}). The models
of soft scattering via polarization of the nonperturbative
QCD vacuum by Nachtmann et al. \cite{Nachtmann,Dosch}
belongs to this category and gives $\sigma_{npt}(r)$ very
similar to our parameterization (\ref{eq:5.1}). 
In the interesting
region of $r\lsim (1--1.5)$fm, the conservative estimate
of uncertainties in $\sigma_{npt}(r)$ is 10-20$\%$, the major
source of uncertainty being due to absorption corrections. For heavy
quarkonia the absorption correction are negligible, though \cite{NNZscan}.

We shall assume the conventional Regge rise of the diffraction
slope for the soft pomeron,
%
%
$
B_{npt}(\xi,r)=\Delta B_{d}(r)+\Delta B_{N}+
2\alpha_{npt}^{'}(\xi-\xi_{0})\,,
$
%
%
where $\Delta B_{d}(r)$ and $\Delta B_{N}$ stand for the contribution
from the beam dipole and target nucleon size and $\xi_{0}=
\log{1\over x_{0}}$. As a guidance we take the experimental
data on the pion-nucleon scattering
\cite{Schiz}, which suggest $\alpha'_{npt}=0.15$\,GeV$^{-2}$
(for the small $\alpha_{npt}'$ descriptions of nucleon-nucleon scattering
see \cite{KNP89}). A plausible guess for the proton size contribution
is
%
%
\beq
\Delta B_{N}=\Delta b_{3}={1\over 3}R_{N}^{2}\, .
\label{eq:5.2}
\eeq
%
%
In the energy
independent soft exchange 
for small dipoles $\Delta B_{d}(r)$ is likely to follow the geometric
law $\Delta B_{d}(r) \approx {1\over 8}r^{2}$ as in Eq.~(\ref{eq:4.9}).
Extension of this law to large dipoles is questionable. 
The large-$r$ saturation of  $\sigma_{npt}(r)$ as parameterized by 
(\ref{eq:5.1}) is a simplifying assumption, what happens at
$r\gg 1$fm is immaterial because even in hadrons the probability of
finding large dipoles, $r\gg 1$fm, is negligible. 
However, the diffraction slope is more 
sensitive to the large dipole contribution.
For instance,
if scattering of large dipoles of size $r\gsim R_{N}$ is modeled
by scattering of thin classical strings off the strongly absorbing
target nucleon of radius $R_{N}$, then for large dipoles (strings),
$r \gsim 2R_{N}$, one readily finds the profile function
%
%
$
\Gamma(b,r)\approx \theta(R-b) + {2\over \pi}
\theta(b-R)
\theta(R+{1\over 2}r - b)
{\rm arcsin}{R \over b}\, ,
$
%
%
which gives the large-$r$ behavior $\sigma_{npt}(r)\approx
2R_{N}r$ and the tamed rise of the diffraction slope
%
%
$
\Delta B_{d}(r\gg 1\,{\rm fm}) \sim {1\over 24}r^{2}\, .
$
%
%
This consideration suggests the interpolation formula of the form
%
%
\beq
B_{d}(r) = {r^{2} \over 8}\cdot
{r^{2}+aR_{N}^{2} \over 3r^{2}+aR_{N}^{2}}\,,
\label{eq:5.3}
\eeq
%
%
where $a$ is a phenomenological parameter, $a\sim 1$.

Such a taming of the growth of $B_{d}(r)$ is supported by the
phenomenology of $\pi N$
scattering.
Take for pion the oscillator wave function and assume that
the gluon probed radius of the pion equals the charge radius.
Then, the contribution from the pion size to the
diffraction slope  for the
purely geometrical form (\ref{eq:4.9}) for $B_{d}(r)$ gives
the unacceptably large
%
$$
 \Delta B_{\pi} ={1\over 8}\cdot
{\langle \pi|
{r^{2}\over 8}
[\sigma_{pt}(x_{0},r)
+\sigma_{npt}(r)]|\pi\rangle \over
 \langle \pi|
{r^{2}\over 8}
[\sigma_{pt}(x_{0},r)
+\sigma_{npt}(r)]|\pi\rangle}
\approx 9.7\,{\rm GeV}^{-2}\, .
\label{eq:5.4}
$$
%
%
Taking for the contribution from the proton size $\Delta B_{N}$ the
estimate 
(\ref{eq:5.2})
we end up with $B_{\pi N}
\approx 15$\,GeV$^{-2}$, which substantially exceeds  the experimental
result $B_{\pi N}(\nu=200\,{\rm GeV}) = 9.9 \pm 0.1\,{\rm GeV}^{-2}$
\cite{Schiz}. The discrepancy increases further if one adds to the
above theoretical estimate the Regge term $2\alpha_{npt}'(\xi-\xi_{0})
\approx 1$\,GeV$^{-2}$ evaluated using the relationship
between the $x_{eff}$ and pion energy
%
%
$
x_{eff} \approx {m_{V}^{2} \over 2\nu m_{p} }\, .
$
%
%

What is the origin of this discrepancy? If $\sigma(\xi,r)$ was $r$ 
independent and if the gluon probed and charge radii of
the pion were identical,
then one would find from (\ref{eq:5.4}) the familiar
%
\beq
\Delta B_{\pi}(\xi_{0}) = {1\over 3}\langle R_{ch}^{2} \rangle_{\pi}
\approx 4\,{\rm GeV}^{-2}\, .
\label{eq:5.5}
\eeq
%
%
With our parameterization (\ref{eq:5.1}), the soft dipole cross section 
keeps rising at $r \sim 1$\,fm and for this reason the matrix element
(\ref{eq:5.4}) is dominated by $r^{2}$ larger than in the
charge radius of the pion and we end up with $\Delta B_{\pi}$
larger than an expectation (\ref{eq:5.5}) based on the charge radius
of the pion. 
The matrix element (\ref{eq:5.4}) can be made smaller
and compatible with the experiment at the expense 
a rapid saturation of the
soft cross section for large dipoles, $\sigma_{npt}(r\gsim 1\,{\rm fm})
\approx \sigma_{tot}(\pi N)$, when one shall recover the estimate
(\ref{eq:5.5}). This solution must be rejected, though, because
it would lead to negligible fluctuations of the soft dipole cross
section in conflict with the experimental data on the diffraction
dissociation of pions, which require
%
%
\beq
{\langle \pi|\sigma^{2}(\nu_{0},r)|\pi\rangle
-
\langle \pi|\sigma(\nu_{0},r)|\pi\rangle^{2}
\over
\langle \pi|\sigma(\nu_{0},r)|\pi\rangle^{2}}
\approx 0.5
\label{eq:5.6}
\eeq
%
%
An attempt to retain the geometrical law and still agree with
the experiment at the expense of taking $\Delta B_{N} \sim 0$
must be rejected too. We believe that the string model suggested
taming of $B_{d}(r)$ Eq.~(\ref{eq:5.3}) is a more acceptable solution.
Hereafter we take $\Delta B_{N}=\Delta b_{3}=4.8\,{\rm GeV}^{-2}$.
Then the pion-nucleon diffraction slope is reproduced with
reasonable values of the parameter $a$ in the formula (\ref{eq:5.3}):
$a=1.2$ for $\alpha'_{npt}=0.1$\,GeV$^{-2}$ and $a=0.9$
for $\alpha'_{npt}=0.15$\,GeV$^{-2}$. Hereafter we shall use the
latter set of parameters. The dependence of the corresponding
diffraction slopes vs. the dipole size $r$ is shown in Fig.~2.
For production of heavy
vector mesons the sensitivity to soft contribution is marginal.

%
%
%
%

\section{Soft-hard decomposition of 
total cross sections for $VN$ scattering}

We start presentation of our results from an evaluation of the vector
meson-nucleon total cross section
%
%
\bea
\sigma_{tot}(VN)
 = {N_{c} \over 2\pi}\int_{0}^{1}{ dz \over z^{2}(1-z)^{2} }
\int d^{2}{\bf{r}}
~\left\{m_{q}^{2} \phi(r,z)^{2}+[z^{2} +(1-z)^{2}]
[\partial_{r}\phi(r,z)]^2 \right\}  \sigma(x_{eff},r) \, .
\label{eq:6.1}
\eea
%
%
For the parameterization of lightcone wave functions $\phi(r,z)$
of vector
mesons see \cite{NNPZ96}. The results for $x_{eff}\leq x_{0}=0.03$
are shown in Fig.~3. The
smaller is the radius of the vector meson $V$ the smaller is the
total cross section $\sigma_{tot}(VN)$, to a crude approximation, 
$\sigma_{tot}(VN) \propto R_{V}^{2}$,
excepting the radial excitations $\phi',\rho'$.

In Fig.~3a we show separately the soft pomeron
contribution to $\sigma_{tot}(VN)$. For the $J/\Psi$ the radius
is large, $R_{J/\Psi}\approx 0.4\,{\rm fm} > R_{c}=0.27\,{\rm fm}$
and the soft contribution is substantial, for the $\Upsilon$
the soft contribution is a small correction to the dominant
perturbative contribution. At subasymptotic energies, the
gBFKL approach predicts steeper rise with energy for smaller
dipoles, cf. Eqs.~(\ref{eq:2.9}), 
the trend which is clearly
seen in Fig.~3a. At asymptotic energies the contribution
from the rising gBFKL cross section takes over for all
channels. In \cite{NZdelta} it has been observed that
for the "magic" radius $r_{\Delta} \sim 0.15\,{\rm fm}\sim
{1\over 2}R_{c}$ the gBFKL color dipole cross section exhibits
the precocious asymptotic energy dependence
$\sigma_{pt}(x,r_{\Delta}) \propto x^{-\Delta_{\Pom}}$.
Because $R_{\Upsilon}\approx 0.18$\,fm is very close
to the "magic" radius $r_{\Delta}$, the predicted energy dependence
of the perturbative contribution to $\sigma_{tot}(\Upsilon N)$
is very close to $\propto W^{2\Delta_{\Pom}}=W^{0.8}$.

The case of the $\Psi'$ is interesting for its large radius
$R_{\Psi'}\approx 0.8$\,fm and large soft contribution.
Because the $\Psi'$ and $\phi^{0}$ have very close
radii, the useful comparison is with $\sigma_{tot}(\phi N)$.
For small $W$  $\sigma_{tot}(\Psi' N)$ of the present paper
is indeed numerically very close to $\sigma_{tot}(\phi^{0} N)$
calculated in \cite{NNPZ96}, but the rise of $\sigma_{tot}(\Psi' N)$
by $\sim 50\%$ from $W\sim 10$\,GeV to $W\sim 500$\,GeV is much
weaker than the rise of $\sigma_{tot}(\phi N)$ by almost
a factor 2 over the same energy range. With our energy
independent $\sigma_{npt}(r)$, the rise of $\sigma_{tot}(\Psi' N)$
is entirely due to the perturbative gBFKL cross section
$\sigma_{pt}(\xi,r)$, which rises with energy more steeply at small $r$.
Although the $\Psi'$ and the $\phi^{0}$ have similar mean square
radii, because of the node effect the relative contribution of small
$r$ for the case of $\Psi'N$ is smaller than for the case of $\phi N$
and this explains the counterintuitive difference of the energy
dependence of the two cross sections.

%
%
%
%

\section{Diffractive production cross sections for the $1S$ states
$J/\Psi$ and $\Upsilon$}

Now we turn to the vector meson production. The strong point about
color dipole factorization equations (\ref{eq:2.11}), (\ref{eq:2.12}), 
(\ref{eq:2.23}), (\ref{eq:2.24})
is that apart from the trivial factors $C_{V}$ and $C_{V}/m_{V}$
the production amplitudes are flavor independent when
considered as a function of the scanning radius $r_{S}$
and/or $Q^{2}+m_{V}^{2}$ \cite{KNNZ93,KNNZ94,NNZscan,NZZslope,NNPZ96}. 
To this end, Eqs.~(\ref{eq:2.11}),(\ref{eq:2.12}), 
(\ref{eq:2.23}), (\ref{eq:2.24}) 
represent the
leading twist terms and the correct twist expansion goes in
powers of $1/(Q^{2}+m_{V}^{2})$ rather than in powers of
$1/Q^{2}$.
For instance, in \cite{NNPZ96} we have shown how the ratio of the
$J/\Psi$ and $\rho$ production cross sections becomes
remarkably constant when the two cross sections are taken at
equal $Q^{2}+m_{V}^{2}$ in contrast to a variation by about
three orders in magnitude when the two cross sections are
compared at equal $Q^{2}$.  For this reason we strongly
advocate the presentation of the experimental data as a
function of the flavor-symmetry restoring variable
$Q^{2}+m_{V}^{2}$ rather than $Q^{2}$ and whenever appropriate
we present our results in terms of this scaling variable.

The soft/hard decomposition of production amplitudes depends on
the relationship between $r_{S}$ and $R_{c}$.  The hard
contribution dominates at $r_{S} \lsim R_{c}$, i.e., if 
%
%
\beq
Q^{2}+m_{V}^{2} \gsim {A^{2} \over R{c}^{2}} \sim 30\,{\rm GeV}^{2}\, ,
\label{eq:7.1}
\eeq
which holds better for the heavier vector mesons and the larger $Q^{2}$.
Our phenomenological soft interaction, as well as other models for 
the soft pomeron \cite{Nachtmann,Dosch}, extends well into 
$r\lsim R_{c}$. Arguably, with better understanding of the
perturbative gBFKL amplitude, one can eventually use the
vector meson production for better fixing the effect of
soft interactions at short distances. In Fig.~4 we show
our decomposition of the production amplitudes into the
hard (perturbative) and soft contribution as a function of $Q^{2}$
for different vector mesons at the typical
HERA energy $W=150$\,GeV. Because for the lighter mesons
the pQCD scale parameter is smaller, $\tau_{L}(\rho^{0})
< \tau_{L}(J/\Psi) < \tau_{L}(\Upsilon)$, the soft contribution 
is somewhat larger for the lighter quarkonia.

Regarding a comparison with the experimental data, the most 
straightforward theoretical predictions are
for the forward production  and we calculate 
$d\sigma/dt|_{t=0}$ and $B(t=0)$. The experimental determination
of these quantities requires extrapolations of $d\sigma/dt$ 
to $t=0$, which is not always possible and one often reports the 
$t$-integrated production cross sections. The principal lesson 
from the high precision $\pi^{\pm} N$ scattering 
experiments is that the diffraction slope $B(t)$ 
depends strongly on the region of $t$ and
for average $\langle t \rangle \sim$
0.1-0.2\,GeV$^{2}$ which dominate the integrated total
cross section, the diffraction slope is smaller than
at $t=0$ by $\sim 1$\,GeV$^{-2}$ \cite{Schiz}. We take these $\pi N$
scattering data for the guidance, and for more direct comparison with
the presently available experimental data instead of the directly
calculated $B(t=0)$ in all the cases we report
%
%
\beq
B=B(t=0)-1\,GeV^{-2}
\label{eq:7.2}
\eeq
%
%
which we also use for the evaluation of the $t$-integrated
production cross section from the theoretically calculated
$d\sigma/dt|_{t=0}$:
%
%
\beq
\sigma(\gamma^{*}\rightarrow V)={1\over B}\cdot
\left.{d\sigma(\gamma^{*}\rightarrow V) \over dt}
\right|_{t=0}\, .
\label{eq:7.3}
\eeq
%
%
The uncertainties in the value of $B$ and with the
evaluation 
(\ref{eq:7.3})
presumably do not exceed
$10\%$ and can be reduced when more accurate data will
become available.

We start presentation of our results and comparison with the
available experimental data with real photoproduction of the
$J/\Psi$ in Fig.~5. 
The agreement with experimental data from the fixed target 
experiments (EMC \cite{EMCPsiQ2}; E516 \cite{E516Psi}; 
E401 \cite{E401Psi}; E687 \cite{E687Psi}) and from the HERA 
experiments (ZEUS \cite{ZEUSPsi,ZEUSPsiW96,ZEUSjp97};
H1 \cite{H1Psi,H1PsiW96,H1PsiJ97}) is good in the both magnitude 
and energy dependence of the cross section.
For the $J/\Psi$ and $Q^{2}=0$ the scanning radius
is still large, $r_{s} \approx 0.4$\,fm and Fig.~4 shows that
at smaller energies $W\sim 15$\,GeV
the soft contribution makes $\sim 50\%$ of the photoproduction
amplitude. Still, it is smaller than in $\sigma_{tot}(J/\Psi\, N)$
and $\sigma(\gamma \rightarrow J/\Psi)$ rises much faster
that $\sigma^{2}_{tot}(J/\Psi\,N)$, which is one of examples of
the failure of the vector dominance model for processes with
heavy quarkonia. Recall that VDM predicts $\sigma(\gamma
\rightarrow J/\Psi)\propto \sigma^{2}_{tot}(J/\Psi\,N)$.

The relationship (\ref{eq:2.12})
(and also (\ref{eq:2.5}))
is to a large extent
the model independent one and predicts the dominance of $\sigma_{L}$ at
large $Q^{2}$.  It is convenient to present
the results for $R=\sigma_{L}/\sigma_{T}$ in the form of the
ratio
%
%
\beq
R_{LT}={m_{V}^{2} \over Q^{2}}\left.{d\sigma_{L}(\gamma^{*}\rightarrow V)
\over d\sigma_{T}(\gamma^{*}\rightarrow V)}\right|_{t=0}
\label{eq:7.4}
\eeq
%
%
shown in Fig.~6. 
The point made in \cite{KNNZ94,NNZscan} and in a 
somewhat different form repeated in \cite{Brodsky}
is that compared to ${\cal M}_{L}$
the transverse amplitude ${\cal M}_{T}$ receives larger contribution
from large $r$ asymmetric end-point configurations with
$z(1-z) \ll 1$. For this reason $R_{LT} < 1$ and is steadily
decreasing with $Q^{2}$. The steeper rise of $\sigma_{pt}(x,r)$
at smaller $r$ makes the end-point contributions less
important at higher energies and $R_{LT}$ rises with energy,
although very weakly. 
The above predictions for $R=d\sigma_{L}/d\sigma_{T}$ must be
tested at $t=0$, the present experimental data on $R$ are
for the $t$ integrated cross sections. In \cite{Ivanov} it
has been argued that at large $t$ rather $\sigma_{T}\gg \sigma_{L}$,
so that the ratio $R$ measured experimentally
for the $t$ integrated cross sections can be somewhat smaller
than our predictions for $t=0$. The calculation of the
$t$-dependence of $R_{LT}$ is an interesting subject on its
own and goes beyond the scope of the present analysis.

In the typical HERA kinematics
the longitudinal polarization of the virtual photon 
$\epsilon \approx 1$ and as our predictions for polarization
unseparated production cross section we present
$\sigma(\gamma^{*}\rightarrow V)=\sigma_{T}(\gamma^{*}
\rightarrow V)+\sigma_{L}(\gamma^{*}\rightarrow V)$.
In Fig.~7 we present our predictions for 
the $J/\Psi$ and $\Upsilon$ production. 
The short hand representation of the same
results is in terms of the exponent of the energy dependence
of the $t$-integrated $\sigma(\gamma^{*}\rightarrow V)
\propto  W^{\delta}=W^{4\Delta_{eff}}$ and/or
$d\sigma/dt|_{t=0} \propto W^{\delta}=W^{4\Delta_{eff}}$. The exponent
$\delta$ for the $t$-integrated cross section is slightly 
smaller because of the shrinkage of the diffraction
cone. The effective intercept $\Delta_{eff}$ depends on the range of
$W$ the fit is made (the more detailed discussion of this issue
is found in \cite{NNZscan}), in Fig.~8 we present our evaluations
for $W=100$\,GeV. For the
sake of completeness, we show on the same plot $\Delta_{eff}$
for light vector mesons evaluated from cross sections calculated
in Ref.~\cite{NNPZ96}. Slight departures from exact flavor
symmetry are due to slight differences in the pQCD scale factors
$\tau(V)$ for different vector mesons. 
The predicted downwards turn of $\Delta_{eff}$ at very large 
$Q^{2}$ is due to the increase of $x_{eff}$ at fixed $W$.
The real photoproduction of $\Upsilon$ offers one of the best 
determinations of the intercept $\Delta_{\Pom}$
of the gBFKL pomeron, because in this case one has the 
the magic scanning 
radius $r_{S} \sim {1\over 2}R_{c}$ and we indeed find  
$\Delta_{eff}\approx \Delta_{\Pom}=0.4$. 
The usual fits to the experimental data are 
of the form $\sigma(\gamma^{*}p\rightarrow Vp)
\propto W^{\delta} = W^{4\Delta_{eff}}$. The evaluated value
of $\delta \sim 0.9$ from Fig.~5 in the range $40 < W < 140\,GeV$
is in good agreement with the value $\delta = 0.92\pm 0.14(stat.)\pm
0.10(syst.)$ extracted from the data on elastic $J/\Psi$ photoproduction 
\cite{ZEUSjp97}. 
Analogous estimation 
of $\delta \sim 0.82$ from Fig.~5 in the range $30 < W < 240\,GeV$
is in good agreement with the value $\delta = 0.77\pm 0.13$ 
presented in \cite{H1PsiJ97}. 
The recent H1 data on elastic virtual photoproduction
of $J/\Psi$ \cite{H1PsiJ97} reported the values of $\delta =
1.07\pm 0.57$ at $Q^{2} = 3.7$\,GeV$^{2}$ and
$1.22\pm 0.52$ at $Q^{2} = 16$\,GeV$^{2}$ in the
energy range $40 < W < 150\,GeV$, which correspond with
our results $\delta = 0.98$ and $\delta = 1.15$,
respectively.

The salient features of the $Q^{2}$ dependence are best seen
when cross sections are plotted as a function of the
flavor symmetry restoring variable $Q^{2}+m_{V}^{2}$, when
the $J/\Psi$ and $\Upsilon$ production cross section exhibit
very similar dependence (Fig.~9). With $R_{LT}\approx 1$ the
theory predicts
%
%
\beq
\left.{d\sigma \over dt}\right|_{t=0} \sim
{1 \over (Q^{2}+m_{V}^{2})^{3}}
G^{2}(x_{eff},\tau(V)\cdot (Q^{2}+m_{V}^{2})) \, .
\label{eq:7.5}
\eeq
%
%
If one fits (\ref{eq:7.5}) to the $(Q^{2}+m_{V}^{2})^{-n}$
behavior and neglects the $Q^{2}$ dependence coming
from the gluons structure function, then $n \approx 3$.
The effect of the gluon structure function is twofold.
At a fixed $x_{eff}$, i.e., when energy changes with
$Q^{2}$ according to $W^{2} = (Q^{2}+m_{V}^{2})/x_{eff}$,
the gluon structure function rises with $Q^{2}$, which
lowers the fitted exponent $n$: $n \lsim 3$. On the
other hand, experimentally one usually studies the $Q^{2}$
dependence at fixed energy $W$, when $x_{eff}=
(Q^{2}+m_{V}^{2})/W^{2}$ increases with $Q^{2}$. Because
the gluon structure decreases towards large $x$, this
induced $Q^{2}$ dependence enhances the exponent $n$.
The exponent $n$ depends on the range of $Q^{2}$ the
fit is performed in. For instance, in the $J/\Psi$
production at a typical HERA energy $W=100$\,GeV 
we predict $n\approx 2.8$ for the semiperturbative
region of $Q^{2}\lsim 10$ GeV$^2$ and $n\approx 3.2$ if
the fit is performed in the region of $15\lsim Q^{2}\lsim
100$\,GeV$^{2}$. We recall that for the $\rho^{0}$
production we found $n\approx 2.4$ $Q^{2}\lsim 10$ GeV$^2$
and $n\approx 3.2$ for $15\lsim Q^{2}\lsim
100$\,GeV$^{2}$ \cite{NNPZ96}. The results for the $\Upsilon$ are similar
to the large-$Q^{2}$ result for the $J/\Psi$. The departures
from the exact flavor symmetry due to $R_{LT}\neq 1$ and slight
flavor dependence of the pQCD scale $\tau(V)$ are marginal
for all the practical purposes.

The experimental data on virtual photoproduction of charmonium
states are still scanty and there are no data yet on the
photoproduction of bottonium. In Fig.~10 we present a
summary of the experimental data on the $J/\Psi$ production
from the fixed target EMC experiment  \cite{EMCPsiQ2} and
the ZEUS \cite{ZEUSPsi,ZEUSPsi97Q2} and
 H1 \cite{H1Psi,H1PsiJ97,H1PsiQ2} experiments
at HERA. The theoretical results are for $W=15$\,GeV appropriate
for the EMC experiment (the dashed curve) and for $W=100$\,GeV
appropriate for the HERA experiments, there is a reasonable
agreement between the theory and experiment. One of the outstanding
experimental problems at large $Q^{2}$ is a separation of elastic
reaction  $\gamma^{*}p\rightarrow V+p$ from the inelastic background
$\gamma^{*}p\rightarrow V+X$ and the low energy EMC data
are well known to have been plagued by the inelastic background.
The contribution from inelastic background to the experimental
cross section may be the reason why we underestimate the
experimental data. One more argument on favor of this point will
be presented in the discussion below of the diffraction slope.

%
%
%
%

\section{Diffraction cone for 
the $V(1S)$ states}

Evidently, the contribution to the diffraction slope
from the $\gamma^{*}$-$V$ transition vertex decreases
with the decreasing scanning radius $r_{S}$, i.e.,
with rising $Q^{2}$ \cite{NZZslope}. At fixed energy
$W$ the value of $x_{eff}$ rises and the rapidity $\xi$
decreases which also diminishes the diffraction slope
because the Regge term becomes smaller
which is an important component of the $Q^{2}$ dependence at
fixed $W$. In this section
we report evaluations of the diffraction slope based on
Eq.~(\ref{eq:3.4}). We use the results of Ref.~\cite{NZZslope}
for the energy and dipole size dependence of $B(\xi,r)$
for gBFKL color dipole amplitude. For the soft pomeron
contribution, we use the parameterizations (\ref{eq:5.1}) and
(\ref{eq:5.3}). According to Fig.~4, the nonperturbative
contribution to the $J/\Psi$ and $\Upsilon$ production
amplitudes is small and our results for the diffraction
slope are insensitive to the soft pomeron effects.
Our definition of the diffraction slope is Eq.~(\ref{eq:7.2})
in section 7 and is meant to correspond to the experimentally
measured slope $B(t)$ at $t\sim$ (0.1-0.15)\,GeV$^{-2}$.

In Fig.~11 we present the predicted energy dependence of
the diffraction slope for real photoproduction of the
$J/\Psi$ and $\Upsilon$. As it was shown in \cite{NZZslope},
at subasymptotic energies the diffraction slope for the
gBFKL amplitude grows rather rapidly, by $\sim 1.4$\,GeV$^{-2}$
when $W$ grows by one order in magnitude from the fixed
target $W=15$\,GeV up to the HERA energy $W=150$\,GeV.
This corresponds to the effective shrinkage rate
$\alpha_{eff}' \approx 0.15$\,GeV$^{-2}$, only at very high
energy beyond the HERA range the  shrinkage will follow
the true slope of the Regge trajectory for the rightmost
gBFKL singularity $\alpha'_{\Pom}=0.07$\,GeV$^{-2}$. The
values of $\alpha'_{\Pom}$ and $\alpha'_{eff}$ are very
sensitive to the gluon propagation radius $R_{c}$ and can
eventually be used to fix this poorly known parameter.
At the moment we explore major properties of the solution
for $R_{c}=0.27$\,fm. In Fig.~11 we also show the diffraction slope
for the $J/\Psi$ production at $Q^{2}=100$\,GeV$^{2}$
which nearly coincides with that for real photoproduction
of the $\Upsilon$. This is still another example of
flavor symmetry restoration, because the scanning radii  
$r_{S}$ for the two reactions are very
close to each other.

The flavor symmetry properties of the diffraction cone can be
seen in  Fig.~12. The curves for
$B(\gamma^{*}\rightarrow V)$ of all the vector
mesons do converge together  as a
function of $Q^{2}+m_{V}^{2}$,
slight departures
from exact flavor symmetry can be attributed to a difference
of the pQCD scale factors $\tau(V)$ for light and heavy vector
mesons.  At fixed $W$, the calculated $Q^{2}$
dependence is an interplay of the changing scanning
radius $r_{S}$ and of the decrease of the Regge component 
with the increase of $x_{eff}$. Fig.~4 shows that for
the light vector mesons and $Q^{2} \lsim 10$\,GeV$^{2}$ the
soft pomeron is substantial and the high precision experimental 
data on the
$\rho^{0},\phi^{0}$ in this region of $Q^{2}$ are indispensable
for better understanding of the soft pomeron. 
In Fig.~12b the same results are presented as a
function of the scanning radius $r_{S}$ as defined by
Eq.~(\ref{eq:1.2}) with $A=6$.  Crude estimates for the
$Q^{2}$ dependence of $B(\gamma^{*}\rightarrow V)$ reported
in \cite{NZZslope} are close to the present results.

We can suggest useful empirical parameterizations for the
diffraction slope.
For production of heavy quarkonia, $V=J/\Psi,\,\Upsilon$, the
$Q^{2}$ dependence of the diffraction slope at $W=100$\,GeV
and in the considered range of $Q^{2}\lsim 500$\,GeV$^{2}$ can
be approximated by
%
%
\beq
B(\gamma^{*}\rightarrow V) \approx \beta_{0} - \beta_{1}
\log\left({Q^{2}+m_{V}^{2} \over m_{J/\Psi}^{2}}\right)
\label{eq:8.1}
\eeq
%
%
with the slope $\beta_{1} \approx 1.1$\,GeV$^{-2}$ and the
constant $\beta_{0}\approx 5.8$\,GeV$^{-2}$. Although
(\ref{eq:8.1}) must be regarded only as a purely empirical
crude parameterization, the logarithmic term (\ref{eq:8.1})
is suggestive of a substantial role of the term 
(\ref{eq:3.29})
in the diffraction slope at high energy. We recall that
the constant $\beta_{0}$ is subject to the choice of the $t$
range, it is value of the slope $\beta_{1}$ which is more
closely related to the gBFKL dynamics. For the light vector
mesons, a somewhat better approximation to the results shown in
Fig.~12 is
%
%
\beq
B(\gamma^{*}\rightarrow V) \approx \beta_{0} - \beta_{1}
\log\left({Q^{2}+m_{V}^{2} \over m_{J/\Psi}^{2}}\right)+
{\beta_{2} \over Q^{2} + m_{V}^{2}}
\label{eq:8.2}
\eeq
%
%
with the same $\beta_{1}=1.1$\,GeV$^{-2}$ as above and with
$\beta_{0}=7.1$\,GeV$^{-2}$, $\beta_{2}=1.6$ for the
$\rho^{0}$ production and
$\beta_{0}=7.0$\,GeV$^{-2}$, $\beta_{2}=1.1$ for the
$\phi^{0}$ production.

The experimental studies of the $Q^{2}$ and energy dependence
of the diffraction slope are in the formative stage. In the
heavy quarkonium sector, only photoproduction of the
$J/\Psi$ has been studied to some extent. 
The experimental
situation is summarized in Fig.~13, at the both fixed
target \cite{E687Psi',E401Psi,NMCPsi} and HERA energy
\cite{ZEUSPsi,ZEUSjp97,H1Psi,H1PsiW96} the error bars are too big for the
definitive conclusions on the presence and/or lack of the
shrinkage of the diffraction cone to be drawn. On the
experimental side, the determinations of the diffraction
slope are very sensitive to the rejection of the inelastic
background. Only the E401 experiment \cite{E401Psi} has
used a technique which allowed a direct selection of the
purely elastic events. The E401 result $B(W=15\,{\rm GeV},Q^{2}=0)
=5.6\pm 1.2$\,GeV$^{-2}$ is consistent with the NMC result
$B(W=15\,{\rm GeV},Q^{2}=0)=5.0\pm 1.1$\,GeV$^{-2}$ \cite{NMCPsi}.
The recent high statistics Fermilab E687 experiment
\cite{E687Psi'} has used the nuclear target and its
determination of the diffraction slope for the quasielastic
scattering, $B(W=20\,{\rm GeV},Q^{2}=0)=7.99\pm 0.23$\,GeV$^{-2}$,
is subject to the model-dependent separation of the coherent
and quasielastic production on nuclei. 
At HERA, the first H1 data
gave $B(W=90\,{\rm GeV},Q^{2}=0)=4.7\pm 1.9$\,GeV$^{-2}$
\cite{H1Psi} and
the first ZEUS data gave $B(W=90\,{\rm GeV},Q^{2}=0)=
4.5\pm 1.4$\,GeV$^{-2}$ \cite{ZEUSPsi}, updated with
 higher statistics to $B(W=90\,{\rm GeV},Q^{2}=0)=
4.6\pm 0.6$\,GeV$^{-2}$ \cite{ZEUSjp97}.   
In 1996 the H1 collaboration \cite{H1PsiW96}
found weak evidence for shrinkage of the diffraction cone:
$B(W\sim 60\,{\rm GeV},Q^{2}=0)=3.7\pm 0.2\pm 0.2$\,GeV$^{-2}$ and
$B(W\sim 120\,{\rm GeV},Q^{2}=0)=4.6\pm 0.3\pm 0.3$\,GeV$^{-2}$.

For virtual production of $J/\Psi$ 
the H1
\cite{H1PsiQ2} reported in 1996 the first data:
$B(W=90\,GeV,<Q^{2}>=18\,{\rm GeV}^{2})=
3.8\pm 1.2(st)^{+2.0}_{-1.6}(syst)$\,GeV$^{-2}$.
Recently the ZEUS collaboration \cite{ZEUSPsi97Q2} presented  
the value of the diffraction slope at $Q^{2}=$\,6 GeV$^{2}$,
$B(W=90\,GeV,<Q^{2}>=6\,{\rm GeV}^{2})=
4.5\pm 0.8(st)\pm 1.0(syst)$\,GeV$^{-2}$.
We predict the
decrease of the diffraction slope from $Q^{2}=0$ to
$Q^{2}=18$\,GeV$^{2}$ by mere $\approx 0.5$\,GeV$^{-2}$,
too small an effect to be seen at the present experimental
accuracy. 

The end-point contribution from asymmetric large size dipoles
with $z(1-z)\ll 1$ is different for the production of the
$T$ and $L$ polarized vector mesons and makes the average
scanning radius somewhat larger in the case of the $T$
polarization. Consequently, one would expect the inequality
of diffraction slopes $B_{T} > B_{L}$ for the polarization
states. This is indeed the case, but the effect shown
in Fig.14 is negligible numerically even for
the charmonium states, because in the nonrelativistic quarkonium
the end-point effects are strongly suppressed. For the
bottonium states the $B_{T}-B_{L}$ is absolutely negligible.

%
%
%
%

\section{What is special about diffractive production
of the radially excited states $V(2S)$?}

In the diffraction production of radially excited
$2S$ states ($\Psi',\Upsilon'$) the keyword is the node
effect. The radial wave function of the $2S$ state
changes the sign at $r\sim R_{V}(1S)$ and there are
cancelations of contributions to the production
amplitude from large size dipoles, $r \gsim R_{V}(1S)$,
and small dipoles, $r\lsim R_{V}(1S)$, which were
noticed for the first time in
\cite{KZ91,KNNZ93}.
Manifestations of the node effect for diffractive
production of light vector mesons off nuclei have been
discussed in \cite{NNZanom,BZNFphi}. 
The detailed
analysis of the forward real and virtual photoproduction
of light $2S$ states ($\rho',\phi'$) at high energies
is presented in \cite{NNPZ96}. The major subject of
the present discussion is new manifestations of the node
for the diffraction cone.

 The cancellation
pattern obviously depends on the relationship between 
$r_{S}$ and position of the node $r_{n}$ which
is close to the radius of the $1S$ state,  $r_{n} \sim R_{V}(1S)$.
If  $r_{S} \ll R_{V}(1S)$,
then the wrong-sign contribution to the production
amplitudes from dipoles with $r \gsim r_{n}$
is small and cancelations are weak (the undercompensation scenario
of Ref. \cite{NNZanom}). If $r_{S}
\gsim R_{V}(1S)$, then the production amplitude can even be
dominated by the wrong-sign contribution from $r$ above
the node position (the overcompensation scenario). For
the heavy quarkonia the scanning radius $r_{S}$ is
sufficiently small even at $Q^{2}=0$ and the undercompensation
scenario is realized.

At fixed target energies,  the node effect is sufficiently strong
suppresses
the ratio $R_{21}(t=0)=d\sigma(\Psi')/d\sigma(J/\Psi)|_{t=0}$
by almost one order in magnitude (Fig.~15). Evidently, the
smaller is the scanning radius the smaller is the large-$r$
contribution and the weaker is the node effect, so that the ratio
$d\sigma(\Psi')/d\sigma(J/\Psi)|_{t=0}$ rises with $Q^{2}$
as shown in Fig.~15. When the node effect is strong which is
the case for the $\Psi'$ at $Q^{2}=0$, 
then even slight variations of the scanning radius $r_{S}$
can change the strength of the node effect substantially.
For this reason one must not be surprised that at fixed target
energies the ratio $d\sigma(\Psi')/d\sigma(J/\Psi)|_{t=0}$
changes with $Q^{2}$ quite rapidly, on a scale of $Q^{2}$
smaller than the natural scale $m_{V}^{2}$.
The predicted energy dependence of
$d\sigma(\Psi')/d\sigma(J/\Psi)|_{t=0}$ derives from the
faster growth with energy of the dipole cross section
for smaller dipoles which also diminishes the node effect.
In Fig.~16 we show in more detail for the HERA energy
$W=100$\,GeV the $Q^{2}$ dependence of the ratio of the
$t$ integrated cross sections $\sigma(2S)/\sigma(1S)$
evaluated using the diffraction slope $B(2S)$ described below.
The predicted $Q^{2}$ and $W$ dependence of the node effect
is sufficiently strong to be observed at HERA.
Because for the heavier $b$ quarks the scanning radius
in comparison to $R_{V}(1S)$ is substantially smaller
than for the charmed quarks, the node effect in the
bottonium production is much weaker, the ratio
$d\sigma(\Upsilon')/d\sigma(\Upsilon)|_{t=0}$ is
larger and exhibits much weaker $Q^{2}$ and $W$
dependence than for the charmonium states (Fig.15).

The node effect is slightly different for the $T$ and
$L$ polarization. This is best seen in Fig.~6 which
shows that the ratio $R_{LT}(2S)$ for the $V'(2S)$
production which is different from $R_{LT}(1S)$ both
in the magnitude and $Q^{2},W$ dependence.

The new effect which we focus here on is a nontrivial
impact of the node effect on the diffraction cone.
In the conventional situation the larger are the radii
of the participating particles the larger is the
diffraction slope and for real photoproduction we
have a clear hierarchy
%
%
\beq
B(\gamma\rightarrow \rho^{0})>B(\gamma\rightarrow \phi^{0})>
B(\gamma\rightarrow J/\Psi)>B(\gamma\rightarrow \Upsilon)
\label{eq:9.1}
\eeq
%
%
which follows the hierarchy of radii $R_{\rho^{0}}>
R_{\phi^{0}}>R_{J/\Psi}>R_{\Upsilon}$.
Although the mean squared radius of the $\Psi'$ is about twice as
large than $R_{J/\Psi}$,, the color dipole approach uniquely
predict $B(\gamma\rightarrow \Psi')<B(\gamma\rightarrow J/\Psi)$
in a striking defiance of the hierarchy (\ref{eq:9.1}), which 
has the following origin: Let ${\cal M}_{<}$ and
${\cal M}_{>}$ be the moduli of contributions to the $V(2S)$
production amplitude from color dipoles with size $r$ below and
above the position of the node in the $2S$ radial wave function
and let $B_{<}$ and $B_{>}$ be the diffraction slopes for the
corresponding contributions. Because of the hierarchy (\ref{eq:9.1})
we have a strong inequality
%
%
\beq
B_{>} > B_{<}\, .
\label{eq:9.2}
\eeq
%
%
For production of the $V(1S)$ state $B(1S)\approx B_{<}$.
Now, the total $V(2S)$ production amplitude equals ${\cal M}(2S)=
{\cal M}_{<}-{\cal M}_{>}$ and for the diffraction slope we
find
%
%
\beq
B(2S)= {B_{<}{\cal M}_{<}- B_{>}{\cal M}_{>} \over
{\cal M}_{<}-{\cal M}_{>}}= B_{<}- (B_{>}-B_{<}){{\cal M}_{>} \over
{\cal M}_{<}-{\cal M}_{>}}\,,
\label{eq:9.3}
\eeq
%
%
which gives an estimate
%
%
\beq
B(2S)-B(1S) \approx
-(B_{>}-B_{<}){{\cal M}_{>} \over
{\cal M}_{<}-{\cal M}_{>}} < 0\,.
\label{eq:9.4}
\eeq
%
%
The weaker is the node effect the smaller is the difference
of diffraction slopes $B(2S)-B(1S)$.
The typical color dipole sizes $r$ which enter $M_{<}$ and $M_{>}$
differ by $\sim R_{V}(1S)$ and the difference of slopes $B_{>}-B_{<}$
can be evaluated as a variation of the diffraction slope $B(1S)$
when the scanning radius $r_{S}$ changes by the factor $\sim 2$
from it's value at $Q^{2}=0$. Then the parameterization gives for
heavy quarkonia an estimate $B_{>}-B_{<} \sim $(1-2)$\beta_{1}
\sim 1 {\rm GeV}^{-2}$. Eq.~(\ref{eq:9.4}) shows the splitting 
$B(2S)-B(1S)$ is further suppressed
if the node effect is weak, i.e., if ${\cal M}_{>} \ll {\cal M}_{<}$.

The results for the $B(1S)-B(2S)$ are presented in Fig.~17. For the
bottonium family the node effect is negligibly weak, cf. Fig.~15, 
whereas for the charmonium family the chances of the
experimental observation of the inequality $B(2S) < B(1S)$ are not
nil at least in real photoproduction and in the fixed target
experiments. The difference of diffraction slopes $B(1S)-B(2S)$
is larger for the $L$ polarization conforming to a stronger node
effect as seen in Fig.~15.
As discussed above and shown in Fig.~15 the node
effect diminishes with energy and the difference of diffraction
slopes $B(1S)-B(2S)$ drops by the factor $\sim 2$ from the fixed
target to HERA energies. It vanishes at large $Q^{2}$ following
the demise of the node effect in Fig.~15, the remarks on the
rapid variation of the node effect on a scale of $Q^{2}$ smaller
than $m_{V}^{2}$ at fixed target energies $W\sim 15$ GeV are
fully relevant to $B(1S)-B(2S)$ too.

Another demonstration of the node effect leading to inequality,
$B(2S) < B(1S)$, is presented in Fig.~18 in the form of
the $t$- dependence of the differential cross section ratio 
$$
R_{V'/V}(t) = 
\frac{d\sigma(\gamma\rightarrow V')/dt} 
{d\sigma(\gamma\rightarrow V)/dt} 
$$
for real photoproduction.
The ratio $R_{\Psi'/(J/\Psi)}(t)$ rises with $t$ at $W = 15$\,GeV as
a consequence of the node effect, whereas at $W = 100$\,GeV
Fig.~18 shows practically constant $t$- dependence of
$R_{\Psi'/(J/\Psi)}(t)$ and $R_{\Upsilon'/\Upsilon}(t)$ 
corresponding to a weaker node effect
at larger energy and for heavier vector mesons, respectively, 
see also Fig.~17.

There is a solid experimental evidence for the node effect in
real photoproduction of the $\Psi'$. In 1996 the H1 collaboration
reported the first observation of real photoproduction of the
$\Psi'$ at HERA with the result 
$R_{21} =\sigma(\gamma \rightarrow \Psi')
/\sigma(\gamma \rightarrow J/\psi)= 0.15\pm 0.05$ \cite{H1Psi'}. 
Concerning the fixed target experiments, only E401 has used the hydrogen
target with the result $\sigma(\gamma \rightarrow \Psi')
/\sigma(\gamma \rightarrow J/\psi)= 0.20\pm 0.05$ at $W=17$\,GeV.
Nuclear targets have been used in all other experiments. Evaluation
of the cross section ratio for the hydrogen target from these
data requires corrections for the nuclear shadowing of the
$J/\Psi$ and nuclear antishadowing of the $\Psi'$ production,
there are also systematic uncertainties with the separation of
coherent and incoherent production.
Specifically, within the same color dipole model as used in
this paper it has been shown \cite{KZ91} that the ratio
$R_{21}=\sigma(\gamma \rightarrow \Psi')/\sigma(\gamma
\rightarrow J/\psi)$ is enhanced in incoherent production off
nuclei by the factor 1.26, 1.55 and 2.16 for the
Be, Fe and Pb nuclei, respectively. For the relatively dilute
$^{6}Li$ target the enhancement factor can be estimated as
$\approx 1.1$. Then, the E687 result $R_{21}(E687)=0.21\pm 0.02$
for the Be target at $W=19$\,GeV \cite{E687Psi'} corresponds
to $R_{21}(E687;N)=0.17\pm0.02$ for the free nucleon target,
the NMC result $R_{21}=0.20\pm 0.05(stat.)\pm 0.07(syst.)$
\cite{NMCPsi'}
after correction for the last value \cite{PDT} of the branching ratio
$BR(J/\Psi\rightarrow \mu^{+}\mu^{-})=5.97\pm 0.25\%$ gives
$R_{21}=0.17\pm 0.04(stat.)\pm 0.04(syst.)$ for the
passive concrete absorber at
$W=15$\,GeV which corresponds to $R_{21}(NMC;N)=0.13\pm 0.05$
for the free nucleon target.
The NA14 result $0.22\pm 0.05$ for the Li target at
$W=14$\,GeV \cite{NA14Psi'}
corresponds to $R_{21}(NA14;N)=0.2\pm0.05$
for the free nucleon target and
the SLAC result $0.22\pm 0.08$ for the Be target at
$W\sim 6.5$\,GeV \cite{SLACPsi'} 
corresponds to $R_{21}(SLAC;N)=0.18\pm 0.07$
for the free nucleon target.
In Fig.~19 we compare our
prediction for $R_{21}(N)=\sigma(\gamma \rightarrow \Psi')
/\sigma(\gamma \rightarrow J/\psi)$ for real photoproduction
on protons with the H1 and E401 data for the proton target
and the above evaluations of $R_{21}(N)$ from the nuclear
target data. The overall agreement between theory and
experiment is satisfactory. In view of the steady collection
of the data at HERA, the higher precision fixed target
data on the hydrogen target are highly desirable to check
unambiguously the predicted rise of $R_{21}(N)$ with
energy.

%
%
%
%

\section{The summary and conclusions}

The major focus of this work has been on
the forward cone for diffractive real and virtual
photoproduction of ground ($1S$) and radially
excited ($2S$) states of heavy quarkonia in the framework
of color dipole running gBFKL approach.
We presented a detailed discussion of the color
dipole factorization for diffractive amplitudes
and of the relevant pQCD factorization scales
with a strong emphasis on restoration of the flavor
symmetry in the variable $Q^{2}+m_{V}^{2}$. We based
our analysis on solutions of the gBFKL equations for
the dipole cross section \cite{NZHERA,NNZscan} and for
the diffraction slope for the color dipole scattering
amplitude \cite{NZZslope}. Starting from the
same dipole cross section which
provides a good quantitative description of the rise
of the proton structure function at small $x$ \cite{NZHERA,NZZRegge},
we found encouraging agreement
with the experimental data on the $Q^{2}$ and energy
dependence of diffractive  $J/\Psi$ production.
 There are many interesting predictions
for the $J/\Psi$ production to be tested, for instance, the
$Q^{2}$ dependence of the effective intercept $\Delta_{eff}$.

A detailed analysis of the energy and $Q^{2}$ dependence of the
diffraction slope  $B(\gamma^{*}\rightarrow V)$ for vector
meson production is presented here for the first time. Of primary
interest is the shrinkage of the diffraction cone which
follows from the finding \cite{NZZslope} that the gBFKL pomeron
is a set of moving poles. We identified different sources of the
$Q^{2}$ dependence of the diffraction slope. Based
on the solution \cite{NZZslope} of the running
gBFKL equation for the
diffraction slope, we presented detailed calculations of
the $Q^{2}$ and $W$ dependence of  $B(\gamma^{*}\rightarrow V)$. 
The present experimental data on $B(\gamma^{*}\rightarrow V)$
for the $J/\Psi$ production are not yet accurate enough to rule in or
rule out our predictions for the shrinkage of the diffraction cone.

Diffractive production of the radially $2S$ mesons ($\Psi',\Upsilon'$)
is a subject on its own. 
The key new feature of production of the $2S$ states
is the node effect, the destructive interference of contributions
to production amplitude from small and large color dipoles because
of the node in the radial wave function of $2S$ radial excitations.
The resulting strong suppression of the $\Psi'$ photoproduction 
agrees with the available experimental data. An interesting
prediction from the color dipole dynamics to be tested is a rise of
the cross section ratio $\sigma(\gamma\rightarrow \Psi')/
\sigma(\gamma \rightarrow J/\Psi)$ by the factor two from the
CENR-FNAL to HERA energies.
The new consequence of the node effect which we discussed
in this paper is a counterintuitive inequality of diffraction slopes
$B(\gamma \rightarrow \Psi') < B(\gamma \rightarrow J/\Psi)$
to be contrasted to a familiar rise of the diffraction slope for
elastic scattering processes with the rise of the radius of the
beam and target particles. The scanning phenomenon allows to
control the node effect varying the scanning radius with $Q^{2}$
and we present the corresponding predictions for the $Q^{2}$
dependence of the cross section ratio $\sigma(\gamma \rightarrow
\Psi')/\sigma(\gamma \rightarrow J/\Psi)$ and of the difference of
diffraction slopes $B(\gamma \rightarrow \Psi') < B(\gamma
\rightarrow J/\Psi)$. The predicted effects for the charmonium
family are within the reach of modern experiments. 
The present analysis of diffractive production of heavy
mesons provides a useful benchmark for future applications
to light vector mesons. The experimental comparison of
virtual and real photoproduction of vector mesons will shed light
on the transition between the soft pomeron exchange which
dominates for the $\rho^{0},\omega^{0},\phi^{0}$ production
at small and moderate $Q^{2}$ to the gBKFL pomeron exchange at
higher $Q^{2}$ and/or heavy vector mesons.
%
\vspace{1.0cm}
\noindent \\

%
%
%
%

{\bf Acknowledgements:}

This work was partly supported by the INTAS grant 93-0239ext.
BGZ and VRZ thank Prof. J.Speth for hospitality at IKP, J\"ulich,
where this work has partly been carried out.
%

%
%
%
%

\pagebreak
{\bf Figure captions:}
\begin{itemize}

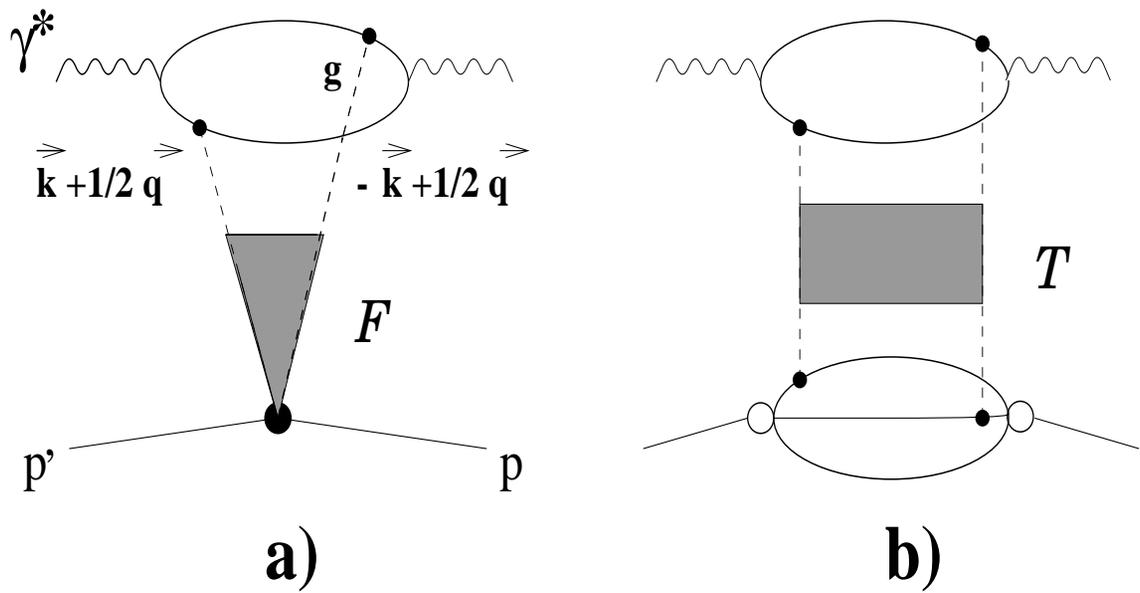
\item[Fig.~1]
~- The perturbative QCD diagrams for vector meson production.

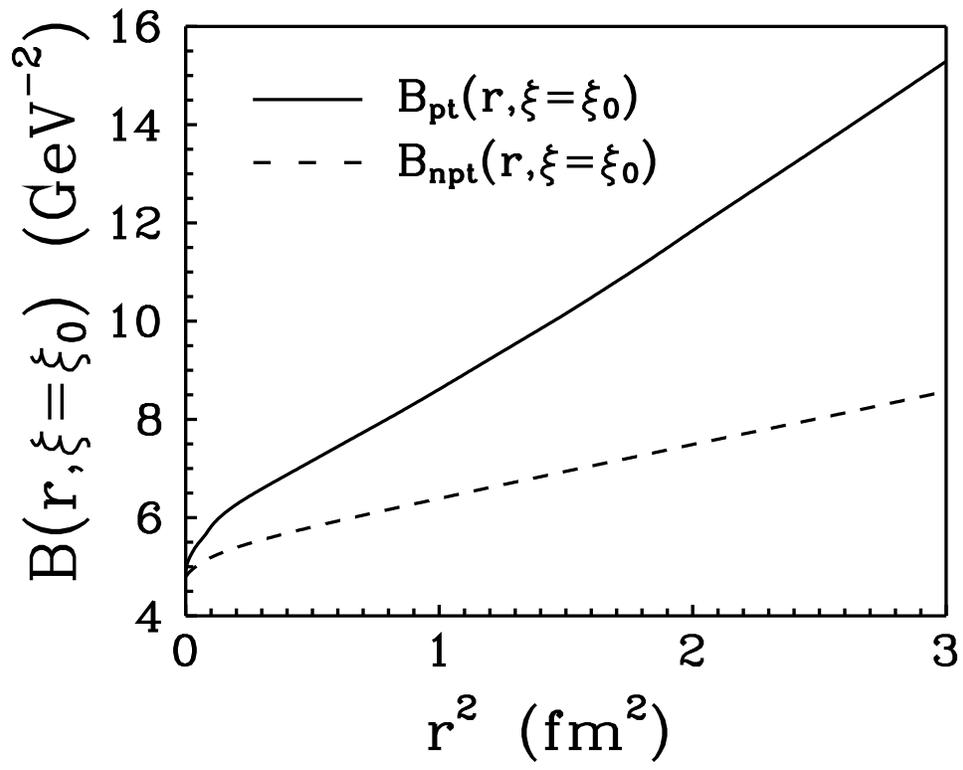
\item[Fig.~2]
~- The dipole size dependence of the diffraction slope 
for the perturbative gBFKL, $B_{pt}(r)$, and soft, $B_{npt}(r)$,
pomerons.

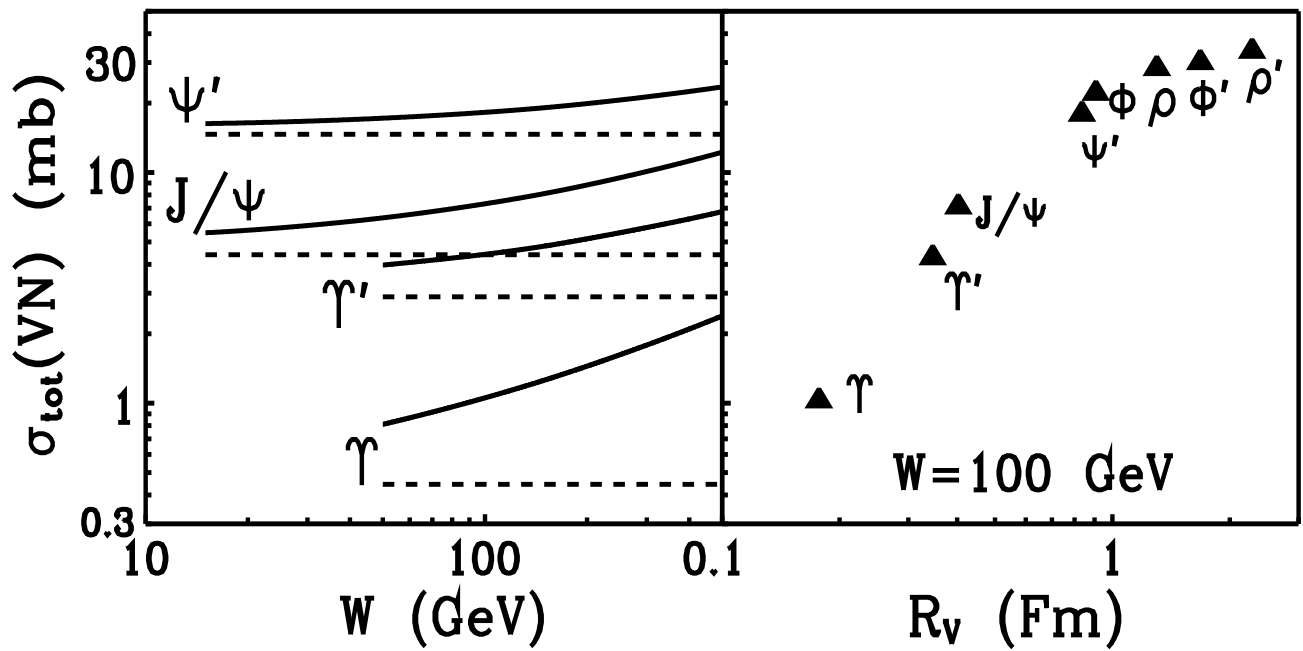
\item[Fig.~3]
~- The color dipole model predictions for
the total cross section $\sigma_{tot}(VN)$
for the interaction of the heavy
vector mesons $J/\Psi, \Psi', \Upsilon$ and $\Upsilon'$ with the nucleon
target as a function of c.m.s. energy $W$.
The dashed shows represent the soft pomeron contribution. 
The right box shows the color dipole model predictions for
the total cross section $\sigma_{tot}(VN)$ vs.the radius $R_{V}$ 
of vector mesons $\rho^{0}$, $\rho'$, $\phi^{0}$, $\phi'$,
$J/\Psi, \Psi', \Upsilon$ and $\Upsilon'$.

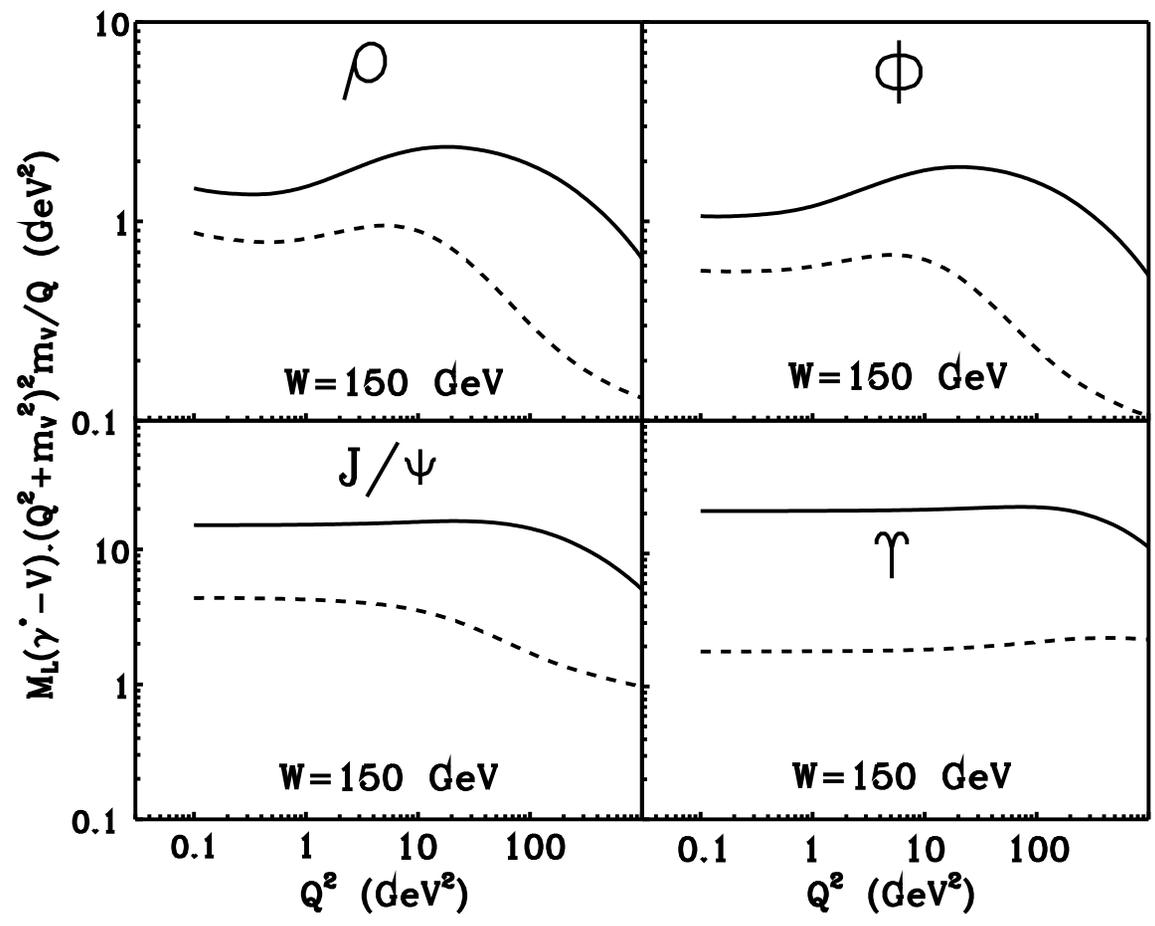
\item[Fig.~4]
~- Decomposition of production amplitude for longitudinally polarized vector
mesons into soft (dashed curves) and perturbative+soft (solid curves)
contribution as a function of $m_{V}^{2}+Q^{2}$. The non-monotonic
$Q^{2}$ dependence is due to the increase of $x_{eff}$ at fixed $W$.

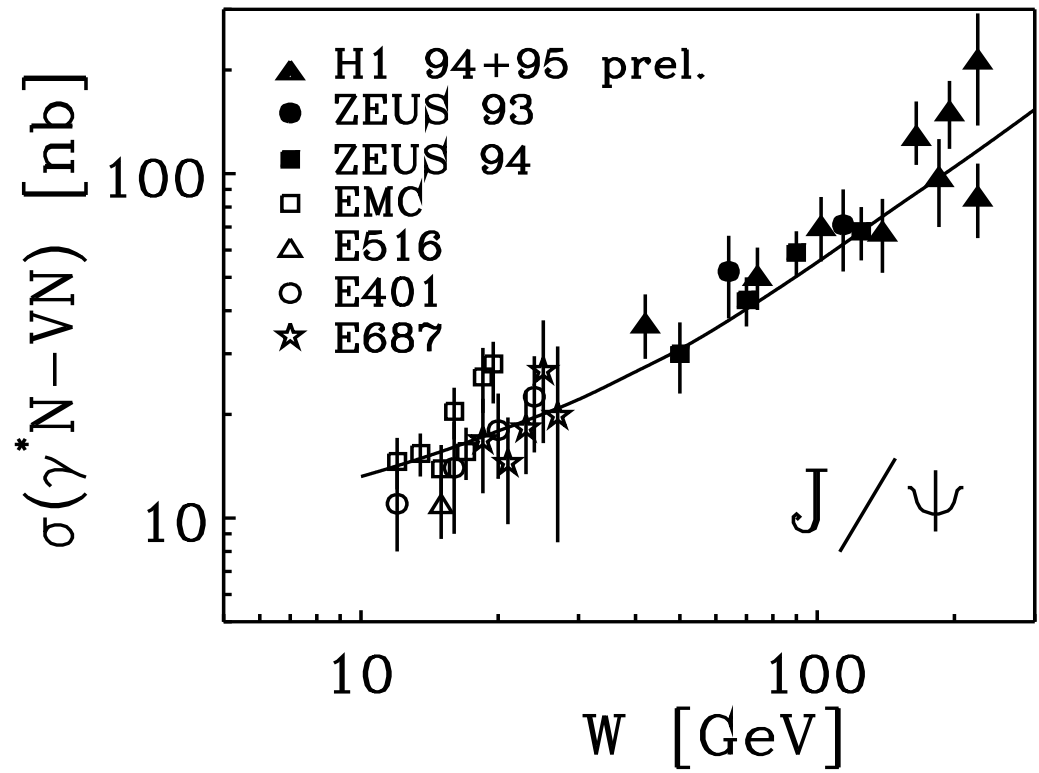
\item[Fig.~5]
~- The color dipole model
predictions for the $W$ dependence of the real photoproduction
cross section $\sigma(\gamma^{*}\rightarrow V)$
for the $J/\Psi$
production vs. the low-energy EMC \cite{EMCPsiQ2},
E516 \cite{E516Psi}, E401 \cite{E401Psi}, E687 \cite{E687Psi}
and high-energy  ZEUS
\cite{ZEUSPsi,ZEUSPsiW96,ZEUSjp97} and H1 \cite{H1Psi,H1PsiW96,H1PsiJ97} data.

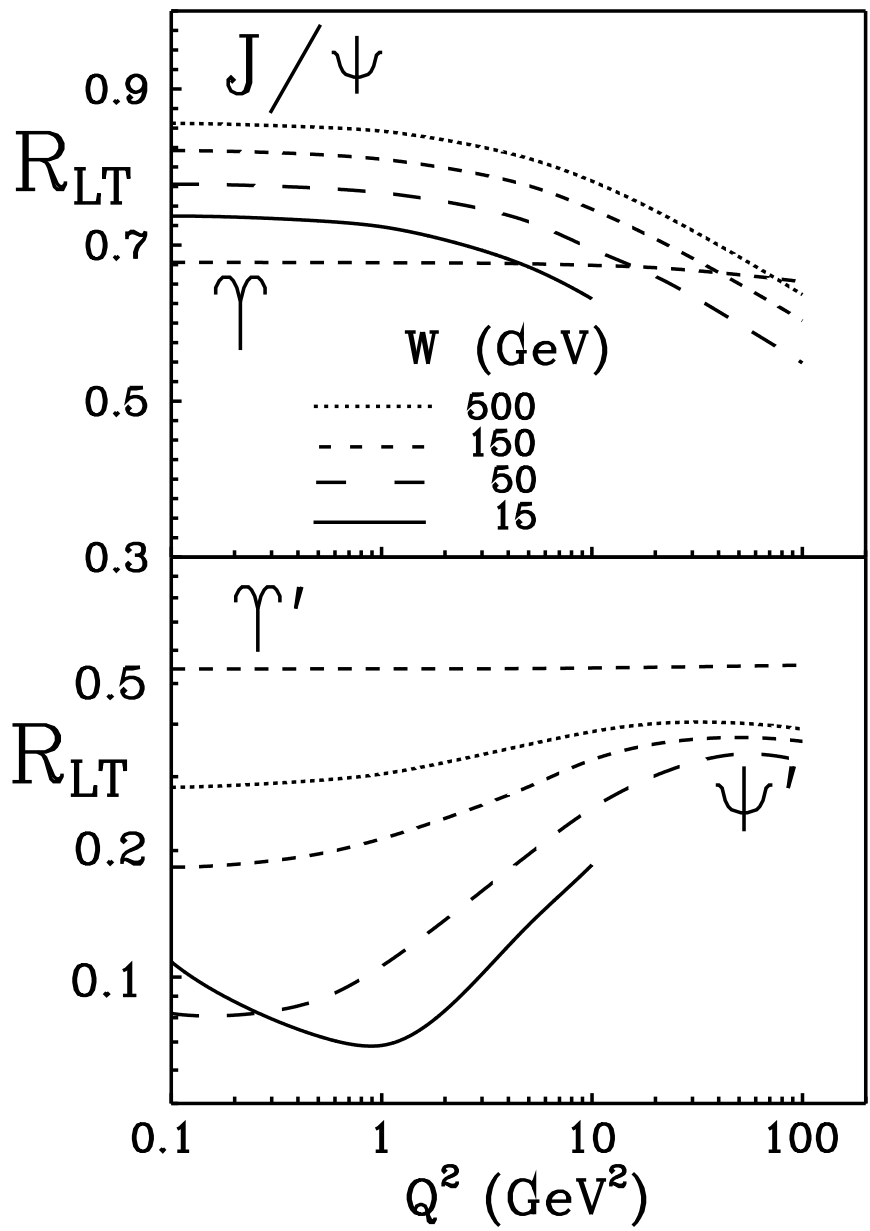
\item[Fig.~6]
~- The color dipole model
predictions for the $Q^2$ and $W$ dependence of the
ratio of the longitudinal and transverse differential cross
sections in the form of the quantity
$
R_{LT}={m_{V}^{2} \over Q^{2}}{d\sigma_{L}(\gamma^{*}\rightarrow V)
\over d\sigma_{T}(\gamma^{*}\rightarrow V)}\,,
$
where $m_{V}$ is the mass of the vector meson.

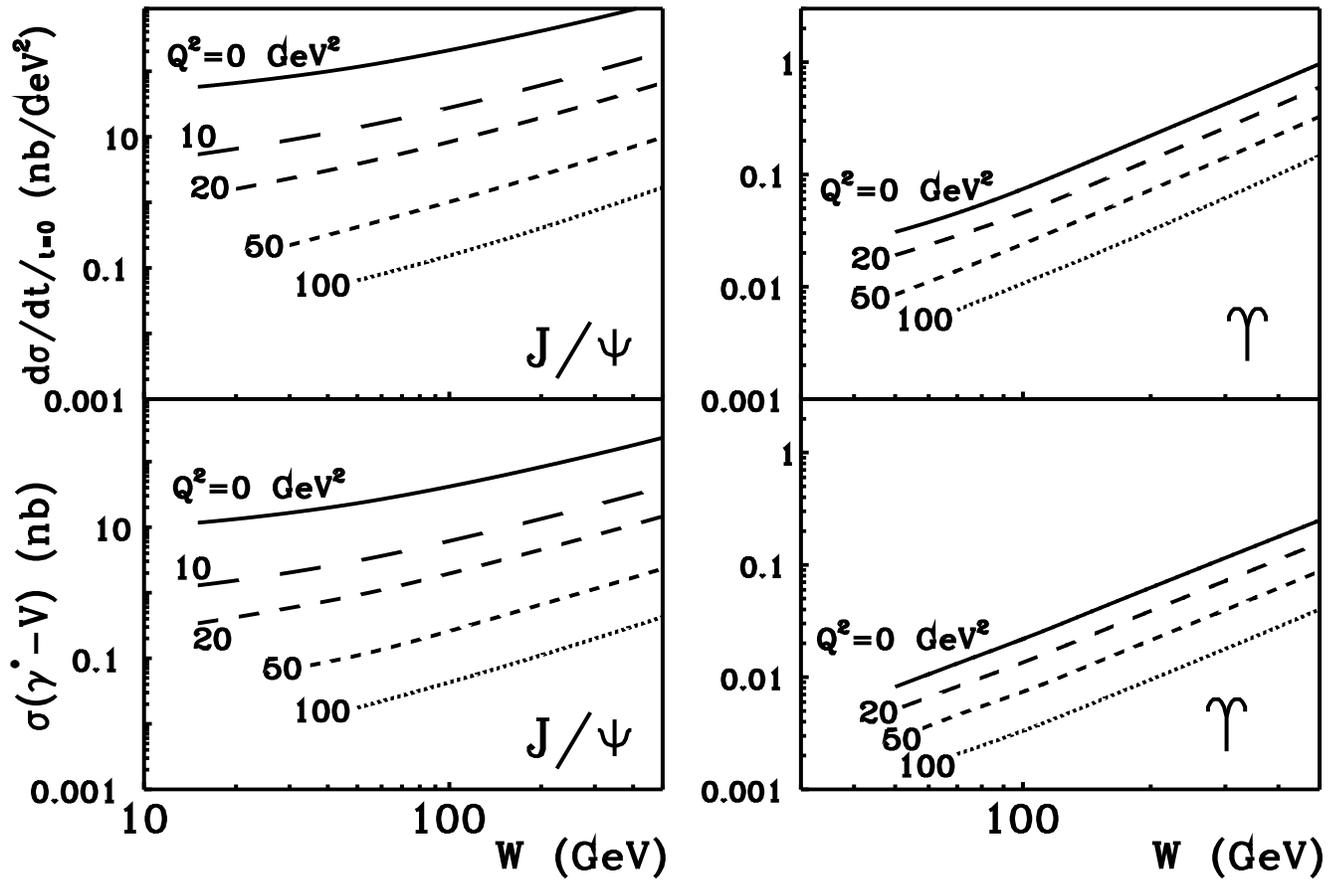
\item[Fig.~7]
~- The color dipole model
predictions for the polarization-unseparated
forward differential cross section (top boxes)
$d\sigma(\gamma^* \rightarrow V)/dt|_{t=0}=
d\sigma_{T}(\gamma^* \rightarrow V)/dt|_{t=0}+
d\sigma_{L}(\gamma^* \rightarrow V)/dt|_{t=0}$
for the $J/\Psi$ and $\Upsilon$ production
as a function of the c.m.s. energy $W$
at different values of $Q^2$. The bottom boxes
are
predictions for the polarization-unseparated
$t$-integrated cross section 
$\sigma(\gamma^* \rightarrow V)=
\sigma_{T}(\gamma^* \rightarrow V)+
\sigma_{L}(\gamma^* \rightarrow V)$.

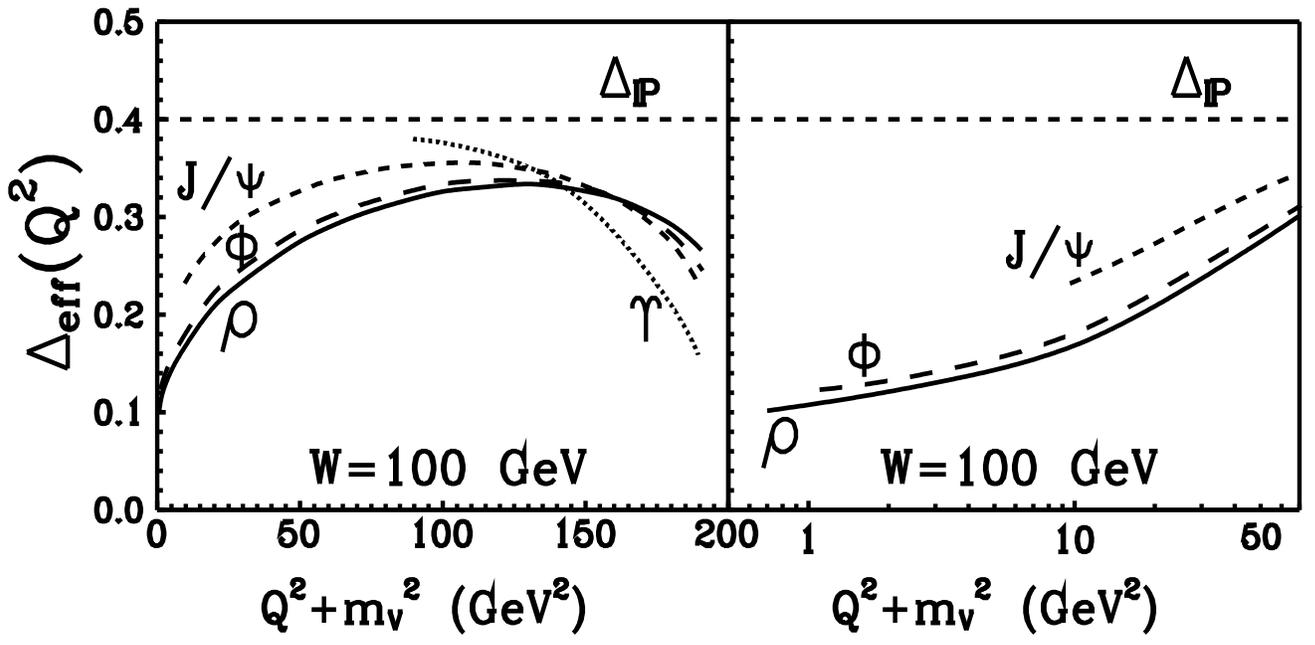
\item[Fig.~8]
~- The $Q^{2}$ dependence of
the effective intercept $\Delta_{eff}(Q^{2})$ for the
forward production of $\rho^{0}$, $\phi^{0}$, $J/\Psi$
and $\Upsilon$.

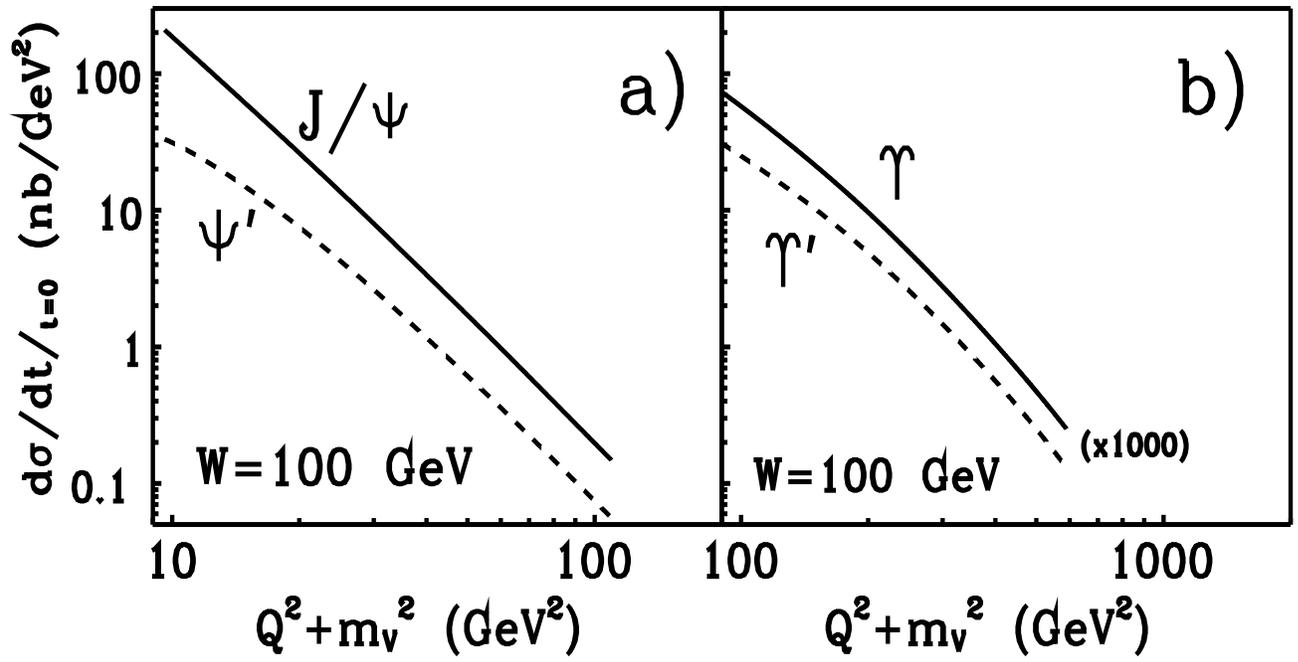
\item[Fig.~9]
~- The color dipole model predictions for the
dependence on the scaling variable
$m_{V}^{2}+Q^{2}$ of the polarization-unseparated
$d\sigma(\gamma^* \rightarrow V)/dt|_{t=0}=
d\sigma_{T}(\gamma^* \rightarrow V)/dt|_{t=0}+
d\sigma_{L}(\gamma^* \rightarrow V)/dt|_{t=0}$
at the HERA energy $W=100\,GeV$.

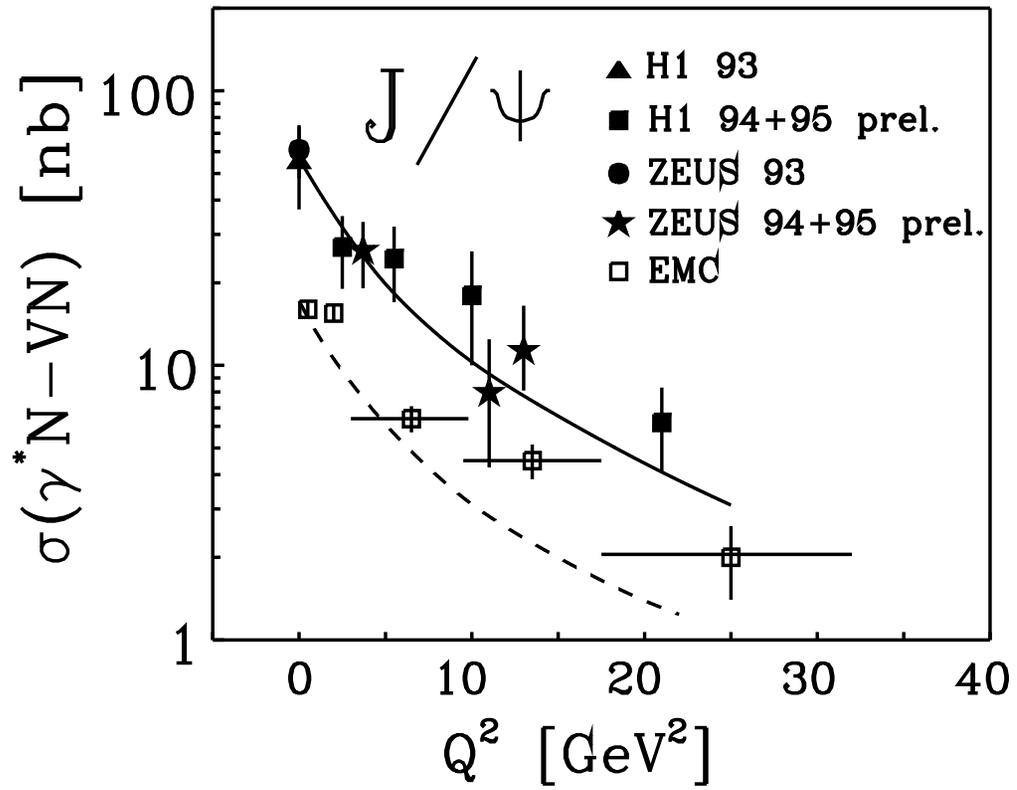
\item[Fig.~10]
~- The color dipole model
predictions for the $Q^{2}$ dependence of the observed
cross section $\sigma(\gamma^{*}\rightarrow V)=
\sigma_{T}(\gamma^{*}\rightarrow V)+\epsilon
\sigma_{L}(\gamma^{*}\rightarrow V)$
of exclusive $J/\Psi$
production vs. the low-energy (EMC \cite{EMCPsiQ2})
and high-energy (ZEUS \cite{ZEUSPsi,ZEUSPsi97Q2}, 
H1 \cite{H1Psi,H1PsiJ97,H1PsiQ2}) data.

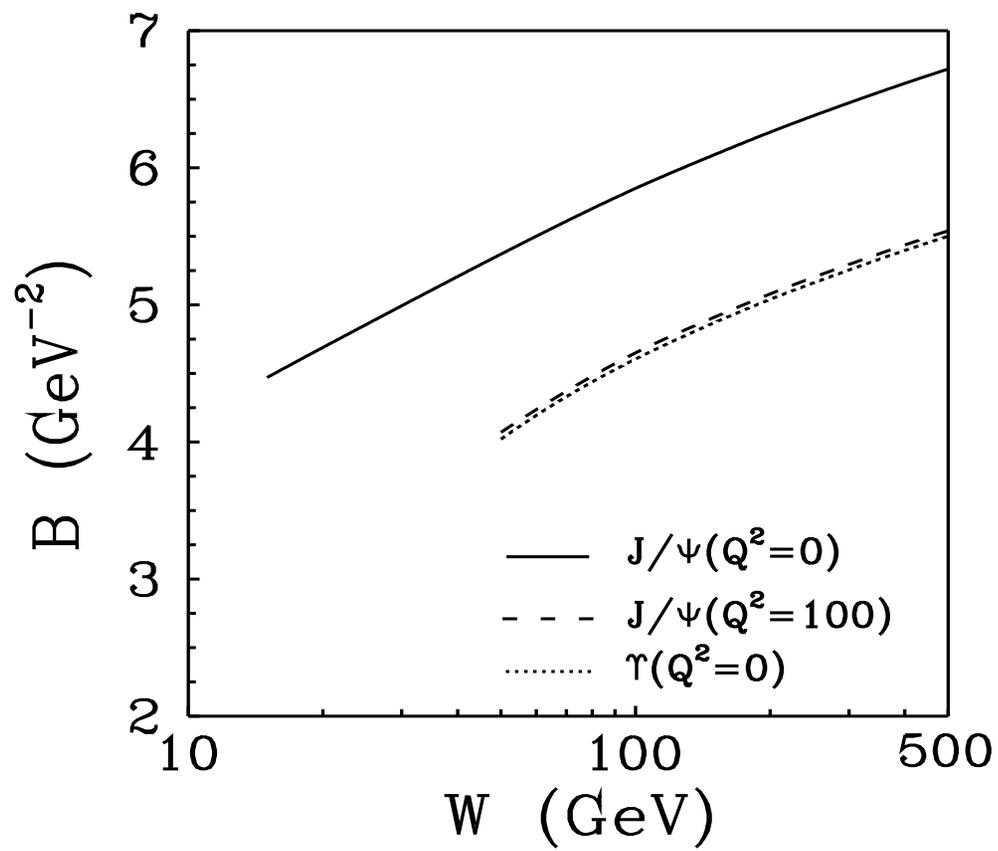
\item[Fig.~11]
~- The color dipole model predictions for the
c.m.s. energy dependence of the diffraction slope
for real photoproduction of the $J/\Psi$ and $\Upsilon$
and for the $J/\Psi$ electroproduction at $Q^{2}=100$\,GeV$^{2}$.

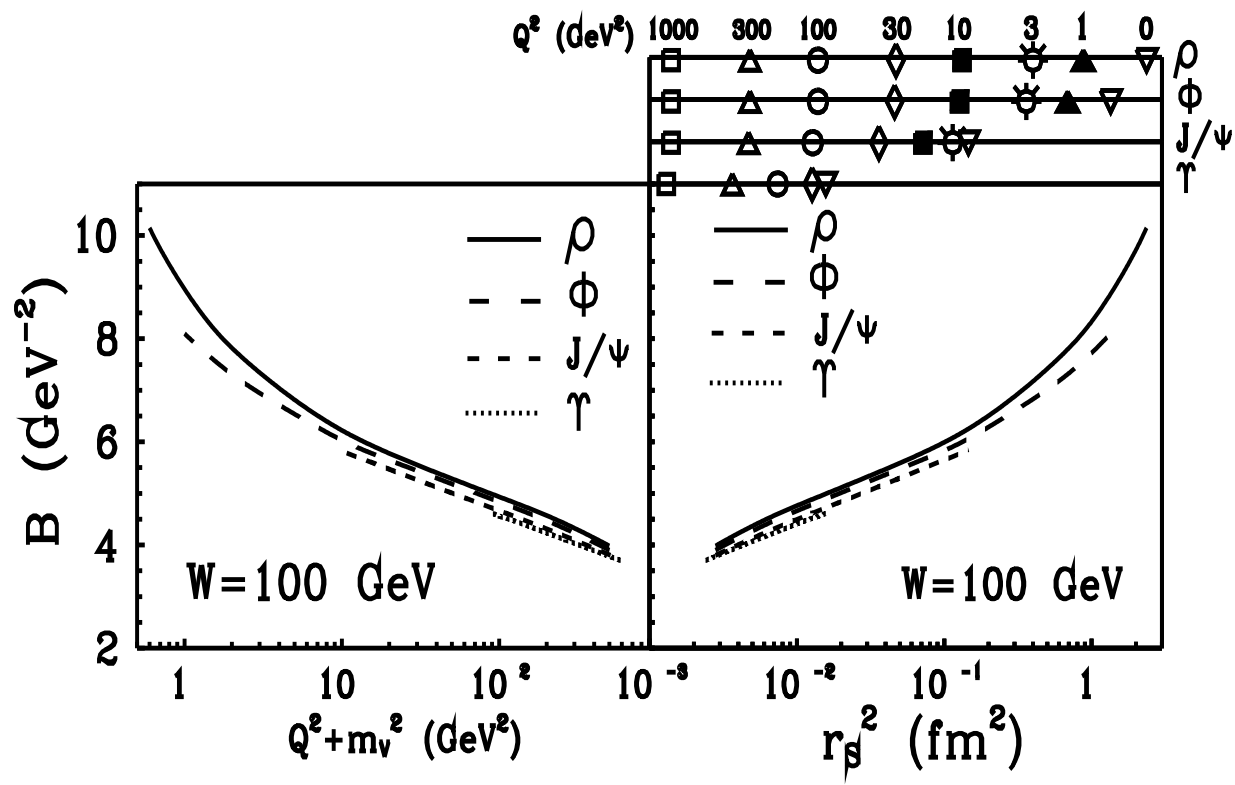
\item[Fig.~12]
~- The color dipole model
predictions for the
diffraction slope in production of different
vector mesons as a function of the scaling variable
$m_{V}^{2}+Q^{2}$ (left box) and the scanning radius $r_{S}$
(right box) at fixed
c.m.s. energy $W=100$\,GeV. 
The scales of $Q^{2}$ on the top of the right box 
show the values of $Q^{2}$ which
correspond to the scanning radii shown on the bottom axis.

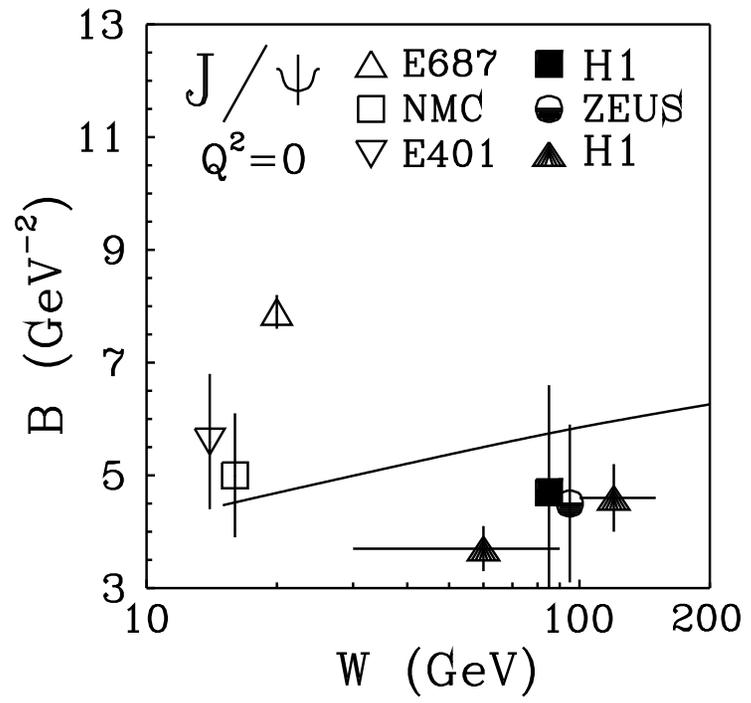
\item[Fig.~13]
~- Comparison of the color dipole model prediction
for c.m.s. energy $W$ dependence of the
diffraction slope for photoproduction of the
$J/\Psi$ with the E401 \cite{E401Psi}, NMC \cite{NMCPsi},
E687 \cite{E687Psi'}, H1 \cite{H1Psi,H1PsiW96} and
ZEUS \cite{ZEUSPsi,ZEUSjp97} data.

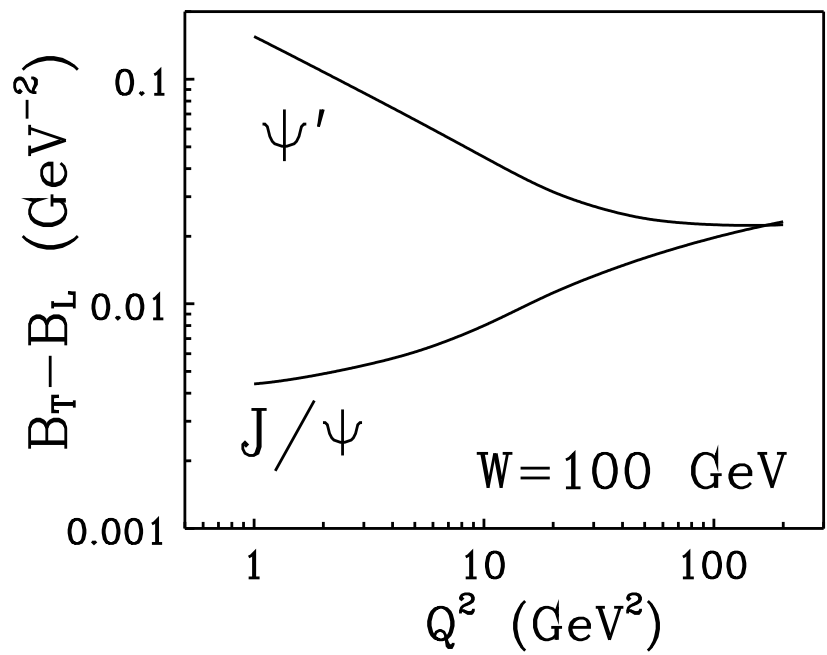
\item[Fig.~14]
~- The color dipole model predictions for the difference
of diffraction slopes $B_{T}-B_{L}$ as a function
of $Q^{2}$ for the $J/\Psi$ and $\Psi'$ production.

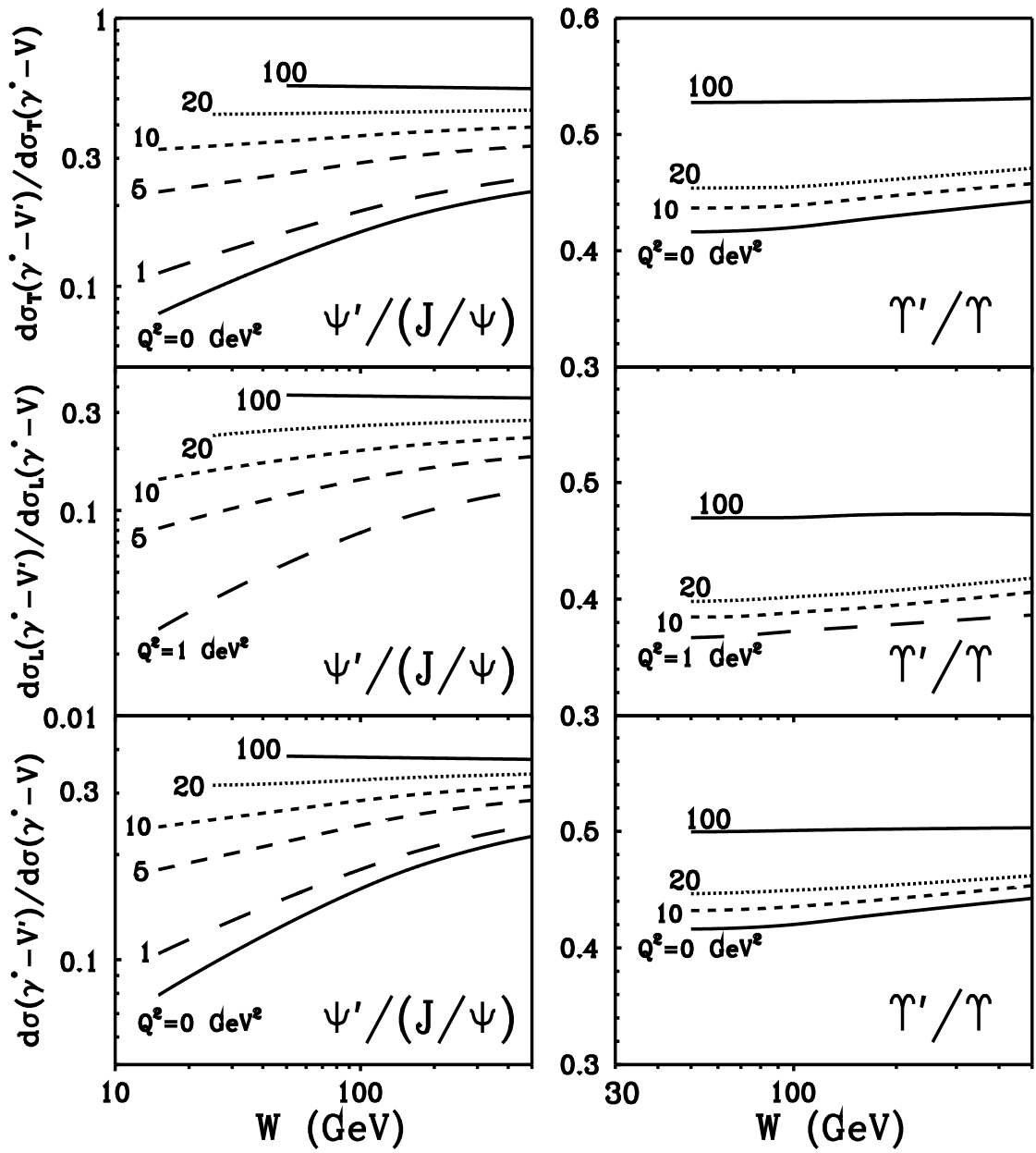
\item[Fig.~15]
~- The color dipole model
predictions for the $Q^2$ and $W$ dependence of the ratios
$d\sigma(\gamma^{*}\rightarrow \Psi'(2S))/
d\sigma(\gamma^{*}\rightarrow J/\Psi)$ and
$d\sigma(\gamma^{*}\rightarrow \Upsilon'(2S))/
d\sigma(\gamma^{*}\rightarrow \Upsilon)$
for the $T$, $L$ and polarization-unseparated 
forward differential cross sections. 

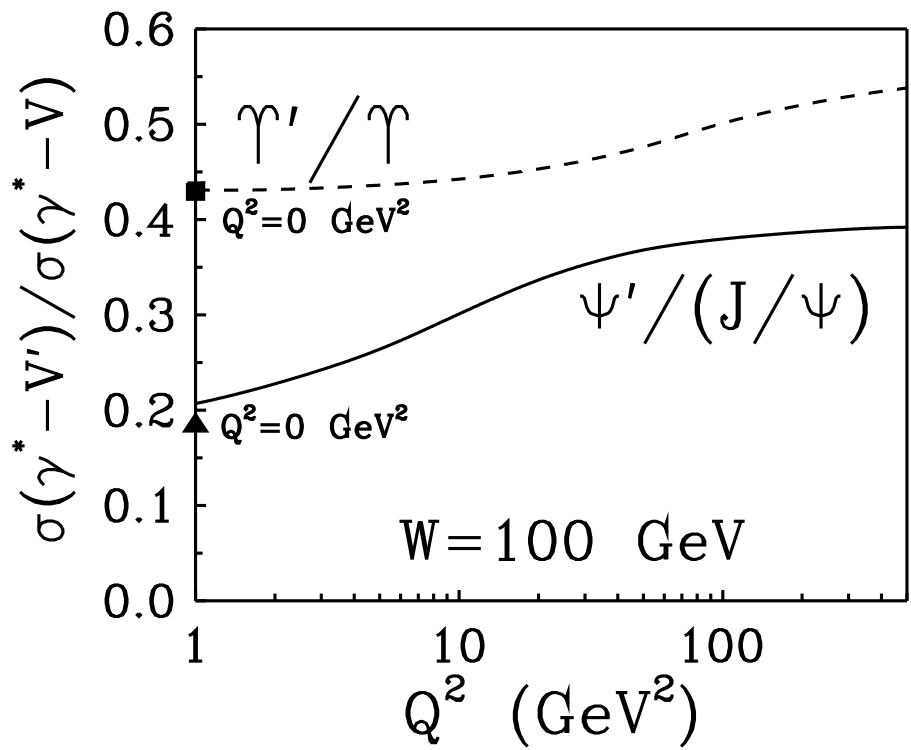
\item[Fig.~16]
~- The color dipole model
predictions for the $Q^{2}$ dependence of the ratio of
the $t$-integrated polarization-unseparated
cross sections
$\sigma(\gamma^{*}\rightarrow \Psi'(2S))/
\sigma(\gamma^{*}\rightarrow J/\Psi)$ and
$\sigma(\gamma^{*}\rightarrow \Upsilon'(2S))/
\sigma(\gamma^{*}\rightarrow \Upsilon)$
 at c.m.s. energy $W=100$
\,GeV.

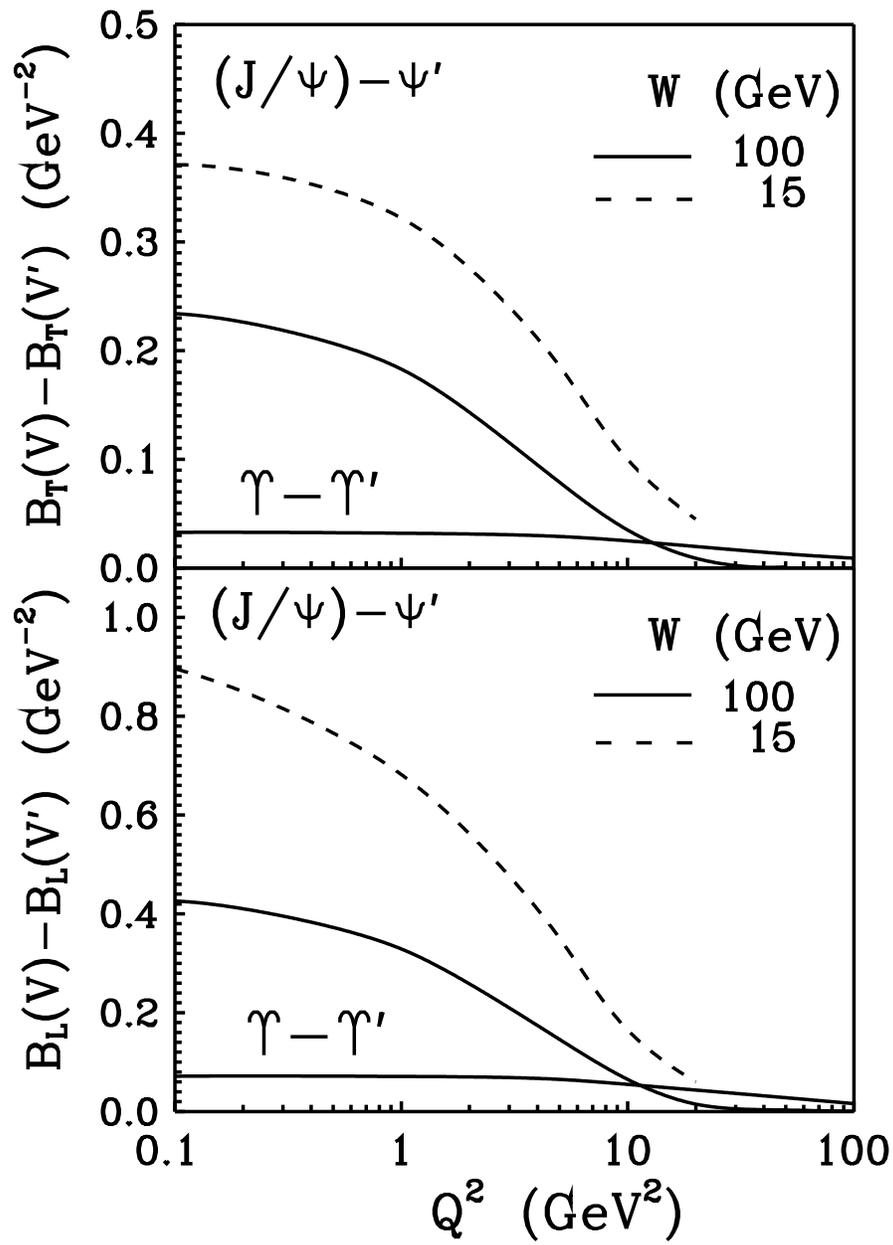
\item[Fig.~17]
~- The color dipole model predictions for
the difference of diffraction slopes $B(1S)-B(2S)$ vs.
$Q^{2}$ at c.m.s. energy $W=15$\,GeV (dashed lines)
and $W=100$\,GeV (solid lines) for $T$ and $L$
polarization.

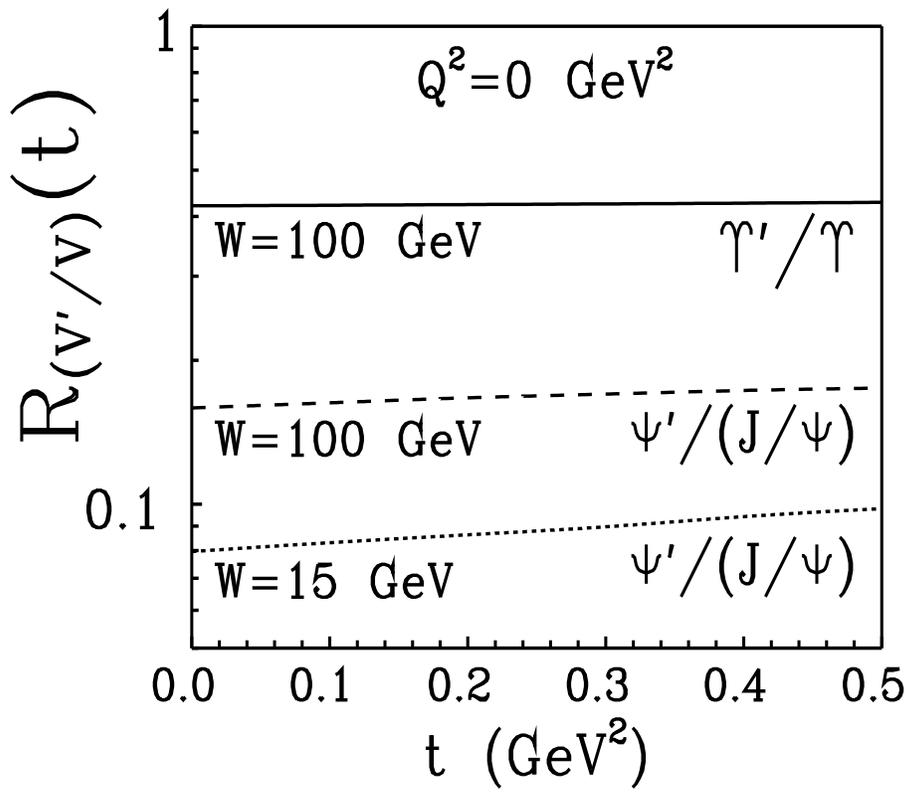
\item[Fig.~18]
~-The color dipole model predictions for
$t$- dependence of
the ratio $R_{V'/V}(t) = {d\sigma(\gamma\rightarrow V')/dt
\over d\sigma(\gamma\rightarrow V)/dt},
$
for the $\Psi'/(J/\Psi)$  and  
$\Upsilon'/\Upsilon$ real photoproduction.

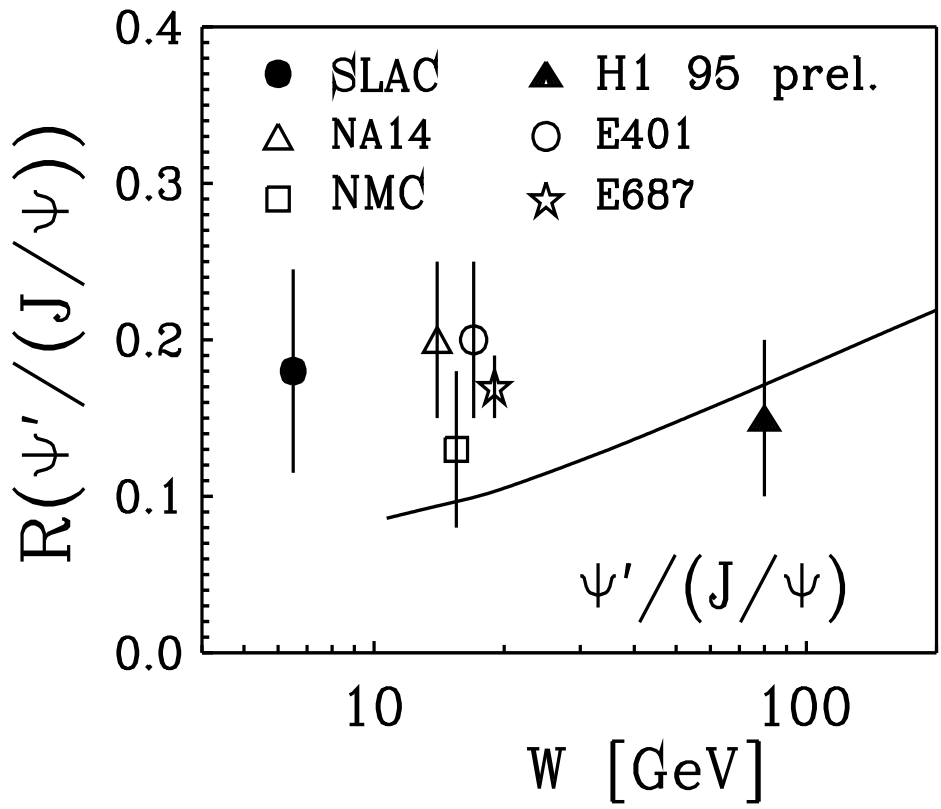
\item[Fig.~19]
~- Comparison of the color dipole model prediction
for c.m.s. energy $W$ dependence of the
ratio $\sigma(\gamma\rightarrow\Psi')/
\sigma(\gamma\rightarrow J/\Psi)$ at $Q^{2}=0$
with the E401 \cite{E401Psi}, NMC \cite{NMCPsi'},
E687 \cite{E687Psi'}, NA14 \cite{NA14Psi'},
SLAC \cite{SLACPsi'} and
H1 \cite{H1Psi'} data.

\end{itemize}

\end{document}